\documentclass[orivec]{llncs}

\frontmatter          

\usepackage{amsmath,amssymb,amsfonts,setspace}
\usepackage{stmaryrd}
\usepackage{enumerate}
\usepackage{cite}

\usepackage{graphicx}
\usepackage{tikz}
\usetikzlibrary{positioning,shapes,shadows,arrows,automata,matrix}
\usetikzlibrary{decorations.pathmorphing,decorations}
\usepackage{tkz-graph}
\usepackage{wrapfig}
\usepackage[ligature,reserved,inference]{semantic}
\usepackage{listings}
\usepackage[pdfborder={0 0 0}]{hyperref}

\usepackage[noend,lined,linesnumbered]{algorithm2e}

\newcommand{\Dd}{\mathcal{D}}

\newcommand{\Ff}{\mathcal{F}}

\newcommand{\Pp}{\mathcal{P}}

\newcommand{\orar}[1]{\vec{#1}}

\newcommand{\sep}{\ast}

\newcommand{\pf}{\rightharpoonup}

\newcommand{\loc}{{\mathbb L}}
\newcommand{\data}{{\mathbb D}}

\newcommand{\lvar}{{\sf LVar}}
\newcommand{\locvar}{{\sf LVar}}

\newcommand{\dvar}{{\sf DVar}}

\newcommand{\vars}{\mathsf{Var}}

\newcommand{\nil}{{\sf nil}}
\newcommand{\dom}{{\sf dom}}

\newcommand{\ldom}{{\sf ldom}}

\newcommand{\slemp}{\mathtt{emp}}

\newcommand{\pbst}{\mathit{bst}}
\newcommand{\pbsthole}{\mathit{bsthole}}
\newcommand{\plseg}{\mathit{lseg}}

\newcommand{\plsegeven}{\mathit{lsegeven}}

\newcommand{\plsegb}{\mathit{lsegb}}

\newcommand{\fnext}{\mathtt{next}}
\newcommand{\fleft}{\mathtt{left}}
\newcommand{\fright}{\mathtt{right}}
\newcommand{\fdata}{\mathtt{data}}

\newcommand{\freev}{\mathtt{free}}

\newcommand{\limp}{\Rightarrow}

\newcommand{\ext}{{\it ext}}

\newcommand{\rSUB}{\textit{SUB}}

\newcommand{\spen}{\textsc{spen}}

\newcommand{\eqmap}{{\tt EQ}}

\newcommand{\proglangkeywordstyle}[1]{\ensuremath{\mathbf{#1}}\xspace}
\reservestyle{\proglangkeyword}{\proglangkeywordstyle}
\newcommand{\true}{\ensuremath{\mathtt{true}}\xspace}

\lstdefinelanguage{program}{%
  basicstyle=\tt,
  classoffset=0,
  keywords={var,const,reg,proc,skip,assume,if,then,else,%
    while,do,call,return,post,await,ewait,yield,init,
    let,in,and,or,true,false,
    for,from,to,
    future, touch,
    fork, rfork, join,
    async, finish,returns,
    spawn, sync, inlet,
    eventloop,
    foreach,
    atomic, method, wait, signal,
    thread, client, begin, end, spawn, repeat, times,
  },
  keywordstyle=\bf,
  classoffset=1,
  morekeywords={g,l},
  keywordstyle=\tt,
  classoffset=0,
  basicstyle=\tt,
  commentstyle=\itshape,
  morecomment=[l]{//},
  morecomment=[s]{/*}{*/},
  morecomment=[n]{(*}{*)},
  mathescape=true,
  escapeinside=`'
}

\lstnewenvironment{program}{%
  \lstset{%
    language={program},
    columns=flexible,
    breaklines=true,
    tabsize=4,
  }
}{}

\lstMakeShortInline"

\begin{document}

\sloppy


\title{On Automated Lemma Generation for Separation Logic with Inductive Definitions\thanks{Zhilin Wu is supported by the NSFC projects (No. 61100062, 61272135, and 61472474), and the visiting researcher program of China Scholarship Council. This work was supported by the ANR project Vecolib (ANR-14-CE28-0018).}}
\author{Constantin Enea\inst{1}, Mihaela Sighireanu\inst{1} and Zhilin Wu\inst{2,1}}
\institute{LIAFA, Universit\'{e} Paris Diderot and CNRS, France \and State Key Laboratory of Computer Science, Institute of Software, \\ \hspace{1cm} Chinese Academy of Sciences, China}
\date{}

\maketitle

\vspace*{-6mm}
\begin{abstract}
Separation Logic with inductive definitions is a well-known approach for deductive verification of programs that manipulate dynamic data structures. Deciding verification conditions in this context is usually based on user-provided lemmas relating the inductive definitions. We propose a novel approach for generating these lemmas automatically which is based on simple syntactic criteria and deterministic strategies for applying them. Our approach focuses on iterative programs, although it can be applied to recursive programs as well, and specifications that describe not only the shape of the data structures, but also their content or their size. Empirically, we find that our approach is powerful enough to deal with sophisticated benchmarks, e.g., iterative procedures for searching, inserting, or deleting elements in sorted lists, binary search tress, red-black trees, and AVL trees, in a very efficient way. 
\end{abstract}


\vspace*{-10mm}

\section{Introduction}\label{sec:intro}

\vspace*{-2mm}

Program verification requires reasoning about complex, unbounded size data structures that may carry data
ranging over infinite domains. Examples of such structures are multi-linked lists, nested lists, trees, etc. Programs
manipulating such structures perform operations that may modify their shape (due to dynamic creation
and destructive updates) as well as the data attached to their elements. An important issue is the design
of logic-based frameworks that express assertions about program configurations (at given control
points), and then to check automatically the validity of these assertions, for all computations. 
This leads to the challenging problem of finding relevant compromises between expressiveness, automation, and scalability.

An established approach for scalability is the use of \emph{Separation logic} (SL)~\cite{ORY01,Rey02}. Indeed, its support for local reasoning based on the ``frame rule''
leads to compact proofs, that can be dealt with in an efficient way. However, finding expressive fragments of SL for writing program assertions,
that enable efficient automated validation of the verification conditions, remains a major issue. 
Typically, SL is used in combination with \emph{inductive definitions}, which provide a natural description of the data structures manipulated by a program.
Moreover, since program proofs themselves are based on induction, using inductive definitions instead of universal quantifiers (like in approaches based on first-order logic)
enables scalable automation, especially for recursive programs which traverse the data structure according to their inductive definition, e.g.,~\cite{QGS+13}.
Nevertheless, automating the validation of the verification conditions generated for {\bf iterative programs}, that traverse the data structures using while loops, remains a challenge.
The loop invariants use inductive definitions for \emph{fragments of data structures}, traversed during a partial execution of the loop, and proving the inductiveness of these invariants requires non-trivial \emph{lemmas} relating (compositions of) such inductive definitions. Most of the existing works require that these lemmas be provided by the user of the verification system, e.g.,~\cite{CHL11,QGS+13,NC08} or they use translations of SL to first-order logic to avoid this problem. However, the latter approaches work only for rather limited fragments~\cite{PWZ13,PWZ14}.  
In general, it is difficult to have lemmas relating complex user-defined inductive predicates that describe not only the shape of the data structures but also their content.


To illustrate this difficulty, consider the simple example of a sorted singly linked list. 
The following inductive definition describes a sorted list segment from the location $E$ to $F$, storing a multiset of values $M$:
%
\vspace{-1eX}
{\small
\begin{align}
\label{rule:lseg1-base}
\plseg(E,M,F) &::= E = F \land M=\emptyset \land \slemp \\ 
\nonumber
\plseg(E,M,F) &::= \exists X,v,M_1.\ E\mapsto\{(\fnext,X),(\fdata,v)\}\sep \plseg(X,M_1,F) \\
\label{rule:lseg1-rec} &\hspace{1.9cm}\land\ v \le M_1\land M=M_1\cup\{v\}
\end{align}
}

\vspace{-4eX}
\noindent
where $\slemp$ denotes the empty heap, $E\mapsto\{(\fnext,X),(\fdata,v)\}$ states that the pointer field $\fnext$ of $E$ points to $X$ while its field $\fdata$ stores the value $v$, and $\sep$ is the separating conjunction. 
Proving inductive invariants of typical sorting procedures requires such an inductive definition and the following lemma:
\vspace{-1eX}
{\small 
\begin{align*}
\exists E_2.\, \plseg(E_1,M_1,E_2) * \plseg(E_2,M_2,E_3) \land M_1\le M_2\limp \exists M.\, \plseg(E_1,M,E_3).
\end{align*}
}

\vspace{-4eX}
\noindent The data constraints in these lemmas, e.g., $M_1\le M_2$ (stating that every element of $M_1$ is less or equal than all the elements of $M_2$), which become more complex when reasoning for instance about binary search trees, are an important obstacle for trying to synthesize them automatically.  

Our work is based on a new class of inductive definitions for describing fragments of data structures that 
(i) supports lemmas {\bf without additional} data constraints like $M_1 \le M_2$ and
(ii) allows to {\bf automatically synthesize} these lemmas using efficiently checkable, almost syntactic, criteria. 
For instance, we use a different inductive definition for $\plseg$, which introduces an additional parameter $M'$ that provides a ``data port'' for appending another sorted list segment, just like $F$ does for the shape of the list segment:
\vspace{-1eX}
{\small
\begin{align}
\label{rule:lseg-base}
\plseg(E,M,F,M') &::= E = F \land M = M' \land \slemp  \\ 
\nonumber 
\plseg(E,M,F,M') &::= \exists X,v,M_1.\ E\mapsto\{(\fnext,X),(\fdata,v)\}\sep \plseg(X,M_1,F,M')\\
\label{rule:lseg-rec}&\hspace{1.9cm}\land\ v \le M_1\land M=M_1\cup\{v\}
\end{align}
}

\vspace{-4eX}
\noindent
The new definition satisfies the following simpler lemma, which avoids the introduction of data constraints:
\vspace{-1eX}
{\small
\begin{align}
\label{lemma:comp-lseg}
\exists E_2,M_2.\, \plseg(E_1,M_1,E_2,M_2) \sep \plseg(E_2,M_2,E_3,M_3) \limp \plseg(E_1,M_1,E_3,M_3).
\end{align}
}

\vspace{-5mm}
\noindent
Besides such ``composition'' lemmas (formally defined in Sec.~\ref{sec:comp}), we define (in Sec.~\ref{sec:derived}) other classes of lemmas needed in program proofs and we provide efficient criteria for generating them automatically. 
Moreover, we propose (in Sec.~\ref{sec:slice}) a proof strategy using such lemmas, based on simple syntactic matchings of spatial atoms (points-to atoms or predicate atoms like $\plseg$) and reductions to SMT solvers for dealing with the data constraints. 
We show experimentally (in Sec.~\ref{sec:exp}) that this proof strategy is powerful enough to deal with sophisticated benchmarks, e.g., the verification conditions generated from the iterative procedures for searching, inserting, or deleting elements in binary search trees, red-black trees, and AVL trees, in a very efficient way. 
The appendix contains the proofs of theorems and additional classes of lemmas.

%
%
%
%
%
%
%
%


\vspace{-4mm}

\section{Motivating Example}
\label{sec:motivation}

\vspace{-2mm}

%

Fig.~\ref{fig:bst_search} lists an iterative implementation of a search procedure for binary search trees (BSTs). 
The property that $E$ points to the root of a BST storing a multiset of values $M$ is expressed by the following inductively-defined predicate:
\vspace{-2mm}
{\small
\begin{align} 
\label{rule:bst-base}
\pbst(E,M) & ::= E = \nil \land M = \emptyset \land \slemp 
\\
\label{rule:bst-rec}
\pbst(E,M) & ::= 
\exists X,Y,M_1,M_2,v.\, E \mapsto \{(\fleft,X), (\fright,Y),(\fdata,v)\}  \\
\nonumber
&\hspace{2.5cm}\sep\ \pbst(X,M_1)\sep \pbst(Y,M_2) \\
\nonumber
&\hspace{2.5cm}\land\ M=\{v\} \cup M_1  \cup M_2 \land M_1 < v < M_2
\end{align}
}

\vspace{-6mm}
\begin{wrapfigure}{l}{0.4\textwidth}
\vspace{-7eX}
{\small
\begin{program}
int search(struct Tree* root, 
            int key) {
  struct Tree *t = root;
  while (t != NULL) {
    if (t->data == key)
      return 1;
    else if (t->data > key)
      t = t->left;
    else 
      t = t->right;
  }
  return 0;
}
\end{program}}
\vspace{-8mm}
\caption{ }
\label{fig:bst_search}
\vspace{-6mm}
\end{wrapfigure}
\noindent 
The predicate $\pbst(E,M)$ is defined by two rules describing empty (eq.~(\ref{rule:bst-base})) and non-empty trees (eq. (\ref{rule:bst-rec})).
The body (right-hand side) of each rule is a conjunction of a pure formula, formed of (dis)equalities between location variables (e.g. $E = \nil$) and data constraints (e.g. $M= \emptyset$), and a spatial formula describing the structure of the heap. 
The data constraints in eq.~(\ref{rule:bst-rec}) define $M$ to be the multiset of values stored in the tree, and state the sortedness property of BSTs. 

The precondition of \texttt{search} is $\pbst(\texttt{root},M_0)$, where $M_0$ is a ghost variable denoting the multiset of values stored in the tree, while its postcondition is  
$\pbst(\texttt{root},M_0) \land (\texttt{key} \in M_0 \rightarrow \mathit{ret} = 1) \land (\texttt{key} \not \in M_0 \rightarrow \mathit{ret} = 0)$, 
where $\mathit{ret}$ denotes the return value.

The while loop traverses the BST in a top-down manner using the pointer variable \texttt{t}. 
This variable decomposes the heap into two domain-disjoint sub-heaps: the tree rooted at \texttt{t}, and the truncated tree rooted at \texttt{root} which contains a ``hole'' at \texttt{t}. 
To specify the invariant of this loop, we define another predicate $\pbsthole(E,M_1,F,M_2)$ describing ``truncated'' BSTs with one hole $F$ as follows:
%
%
%
\vspace{-2mm}
{\small
\begin{align} 
\label{rule:bsth-base}
\pbsthole(E,M_1,F,M_2) & ::= E = F \land M_1 = M_2 \land \slemp \\
\nonumber
\pbsthole(E,M_1,F,M_2) & ::= 
\exists X,Y,M_3,M_4,v.\, E \mapsto \{(\fleft,X), (\fright,Y),(\fdata,v)\} \\
\label{rule:bsth-recr}
&\hspace{2.2cm}\sep\ \pbst(X,M_3) \sep \pbsthole(Y,M_4,F,M_2) \\\nonumber
&\hspace{2.2cm}\land\ M_1=\{v\} \cup M_3 \cup M_4 \land M_3 < v < M_4 
\\
\nonumber
\pbsthole(E,M_1,F,M_2) & ::= 
\exists X,Y,M_3,M_4, v.\, E \mapsto \{(\fleft,X),(\fright,Y),(\fdata,v)\} \\\label{rule:bsth-recl}
&\hspace{2.2cm}\sep\ \pbsthole(X,M_3,F,M_2) \sep \pbst(Y,M_4) \\\nonumber
&\hspace{2.2cm}\land\ M_1=\{v\} \cup M_3 \cup M_4 \land M_3 < v < M_4
\end{align}
}

\vspace{-6mm}
\noindent
Intuitively, the parameter $M_2$, interpreted as a multiset of values, is used to specify that the structure described by $\pbsthole(E,M_1,F,M_2)$ could be extended with a BST rooted at $F$ and storing the values in $M_2$, to obtain a BST rooted at $E$ and storing the values in $M_1$.
Thus, the parameter $M_1$ of $\pbsthole$ is the union of $M_2$ with the multiset of values stored in the truncated BST represented by $\pbsthole(E,M_1,F,M_2)$.
%

Using $\pbsthole$, we obtain a succinct specification of the loop invariant:
\vspace{-2mm}
{\small
\begin{align}
\label{eq:inv-search}
\textit{Inv} ::= \exists M_1.\ \pbsthole(\texttt{root},M_0,\texttt{t},M_1) \sep \pbst({\tt t},M_1) \land (\texttt{key} \in M_0 \Leftrightarrow \texttt{key} \in M_1).
\end{align}
}
%
%

\vspace{-6mm}
\noindent
We illustrate that such inductive definitions are appropriate for automated reasoning, by taking the following branch of the loop:
{\tt assume(t != NULL);} {\tt assume(t->data > key);} {\tt t}$'$ {\tt = t->left}
(as usual, {\tt if} statements are transformed into {\tt assume} statements and primed variables are introduced in assignments).
The postcondition of ${\it Inv}$ w.r.t. this branch, denoted ${\it post}({\it Inv})$, is computed as usual by unfolding the $\pbst$ predicate: 
%
\vspace{-2mm}
{\small
\begin{align}
\nonumber
\exists M_1, Y, v ,M_2, M_3.\, & 
\pbsthole({\tt root},M_0,{\tt t},M_1) \sep  
{\tt t}\mapsto \{(\fleft,{\tt t}'), (\fright, Y), (\fdata,v)\}  
\\\nonumber
&\hfill\sep\ \pbst({\tt t}',M_2) \sep \pbst(Y,M_3) 
\land M_1 = \{v\} \cup M_2 \cup M_3 \land M_2 < v < M_3 
\\\label{eq:post-Inv-search}
&\hfill\land\  ({\tt key} \in M_0 \Leftrightarrow {\tt key} \in M_1) \land  v > {\tt key}.
\end{align}
}

\vspace{-6mm}
\noindent
The preservation of ${\it Inv}$ by this branch is expressed by the entailment ${\it post}({\it Inv})\limp {\it Inv}'$, where ${\it Inv}'$ is obtained from ${\it Inv}$ by replacing ${\tt t}$ with ${\tt t}'$.
%
%

Based on the lemmas, this paper also proposes a deterministic proof strategy for proving the validity of entailments of the form $\varphi_1\limp\exists \vec{X}.\varphi_2$, where $\varphi_1,\varphi_2$ are quantifier-free and $\vec{X}$ contains only data variables\footnote{The existential quantifiers in $\varphi_1$ are removed using skolemization.}. 
The strategy comprises two steps:
(i) enumerating spatial atoms $A$ from $\varphi_2$, and for each of them, carving out a sub-formula $\varphi_A$ of $\varphi_1$ that entails $A$, where it is required that these subformulas do not share spatial atoms (due to the semantics of separation conjunction), and 
(ii) proving that the data constraints from $\varphi_A$ imply those from $\varphi_2$  (using SMT solvers). 
The step (i) may generate constraints on the variables in $\varphi_A$ and $\varphi_2$ that are used in step (ii).
If the step (ii) succeeds, 
then the entailment holds.

For instance, by applying this strategy to the entailment ${\it post}({\it Inv})\limp {\it Inv}'$ above, we obtain
two goals for step (i) which consist in computing two sub-formulas of ${\it post}({\it Inv})$ that entail 
$\exists M_1'.\ \pbsthole({\tt root},M_0,{\tt t}',M_1')$ and respectively, $\exists M_1''.\ \pbst({\tt t}',M_1'')$. 
This renaming of existential variables requires adding the equality $M_1=M_1'=M_1''$ to ${\it Inv}'$.
The second goal, for $\exists M_1''.\ \pbst({\tt t}',M_1'')$, is solved easily since this atom almost matches the sub-formula $\pbst({\tt t}',M_2)$. This matching generates the constraint $M_1''=M_2$, which provides an instantiation of the existential variable $M_1''$ useful in proving the entailment between the data constraints in step (ii).

Computing a sub-formula that entails $\exists M_1'.\ \pbsthole({\tt root},M_0,{\tt t}',M_1')$ requires a non-trivial lemma. Thus, according to the syntactic criteria defined in Sec.~\ref{sec:comp}, the predicate $\pbsthole$ enjoys the following \emph{composition lemma}:
\vspace{-1mm}
{\small
\begin{align}
\label{eq:bsth-lemma}
\big(\exists F,M.\ \pbsthole({\tt root},M_0, F, M)\ \sep  &\ \pbsthole(F,M,{\tt t}',M_1')\big) 
\\\nonumber
&\hfill\limp \pbsthole({\tt root},M_0,{\tt t}',M_1').
\end{align}
}

\vspace{-6mm}
\noindent
Intuitively, this lemma states that composing two heap structures described by $\pbsthole$ results in a structure that satisfies the same predicate. The particular relation between the arguments of the  predicate atoms in the left-hand side is motivated by the fact that the parameters $F$ and $M$ are supposed to represent ``ports'' for composing $\pbsthole({\tt root},M_0,F, M)$ with some other similar heap structures. This property of $F$ and $M$ is characterized syntactically by the fact that, roughly, $F$ (resp. $M$) occurs only once in the body of each inductive rule of $\pbsthole$, and $F$ (resp. $M$) occurs only in an equality with ${\tt root}$ (resp. $M_0$) in the base rule (we are referring to the rules (\ref{rule:bsth-base})--(\ref{rule:bsth-recl}) with the parameters of $\pbsthole$ substituted by $({\tt root},M_0,F, M)$).

Therefore, the first goal reduces to finding a sub-formula of ${\it post}({\it Inv})$ that implies the premise of (\ref{eq:bsth-lemma}) where $M_1'$ remains existentially-quantified. Recursively, we apply the same strategy of enumerating spatial atoms and finding sub-formulas that entail them. However, we are relying on the fact that all the existential variables denoting the root locations of spatial atoms in the premise of the lemma, e.g., $F$ in lemma (\ref{eq:bsth-lemma}), occur as arguments in the only spatial atom of the conclusion whose root location is the same as that of the consequent, i.e., $\pbsthole({\tt root},M_0, F, M)$ in lemma (\ref{eq:bsth-lemma}).
Therefore, the first sub-goal, $\exists F,M.\  \pbsthole({\tt root},M_0, F, M)$ matches the atom $\pbsthole({\tt root},M_0,{\tt t},M_1)$, under the constraint $F={\tt t}\land M=M_1$. This constraint is used in solving the second sub-goal, which now becomes $\exists M_1'.\ \pbsthole({\tt t},M_1,{\tt t}',M_1')$.


The second sub-goal is proved by unfolding $\pbsthole$ twice, using first the rule (\ref{rule:bsth-recl}) and then the rule (\ref{rule:bsth-base}), 
and by matching the resulting spatial atoms with those in ${\it post}({\it Inv})$ one by one. 
Assuming that the existential variable $M_1$ from ${\it Inv}'$ is instantiated with $M_2$ from ${\it post}({\it Inv})$ (fact automatically deduced in the first step), the data constraints in ${\it post}({\it Inv})$ entail those in ${\it Inv}'$. This completes the proof of ${\it post}({\it Inv})\limp {\it Inv}'$.

\vspace{-4mm}

\section{Separation Logic with Inductive Definitions}
\label{sec:logic}

\vspace{-2mm}

Let $\locvar$ be a set of \emph{location variables}, interpreted as heap locations, and $\dvar$ a set
of \emph{data variables}, interpreted as data values stored in the heap, (multi)sets of values, etc. 
In addition, let $\vars=\locvar \cup \dvar$. The domain of heap locations is denoted by $\loc$ while the domain 
of data values stored in the heap
is generically denoted by $\data$. Let $\Ff$ be a set of pointer fields, interpreted as functions $\loc\pf \loc$, and 
$\Dd$ a set of data fields, interpreted as functions $\loc\pf \data$.
The syntax of the Separation Logic fragment considered in this paper is defined in Tab.~\ref{tab:syntaxSL}.

Formulas are interpreted over pairs $(s,h)$ formed of a \emph{stack} $s$ and a \emph{heap} $h$.
The stack $s$ is a function giving values to a finite set of variables (location or data variables) while the heap $h$ is 
a function mapping a finite set of pairs $(\ell,{\it pf})$,
where $\ell$ is a location and ${\it pf}$ is a pointer field, to locations, and a finite set of pairs $(\ell,{\it df})$,
where ${\it df}$ is a data field, to values in $\data$. In addition, $h$ satisfies the condition that for each $\ell \in \loc$, if $(\ell,{\it df}) \in \dom(h)$ for some ${\it df} \in \Dd$, then $(\ell,{\it pf}) \in \dom(h)$ for some ${\it pf} \in \Ff$.
Let $\dom(h)$ denote the domain of $h$, and $\ldom(h)$ denote the set of $\ell \in \loc$ such that $(\ell, {\it pf}) \in \dom(h)$ for some ${\it pf} \in \Ff$.

\begin{table}[t]
\caption{The syntax of the Separation Logic fragment}
\label{tab:syntaxSL}
\vspace{-4eX}
\small
$$
\begin{array}{l}
\begin{array}{ll}
X,Y,E \in \locvar \mbox{ location variables} &  \rho\ \subseteq\ (\Ff\times \locvar)\cup (\Dd\times\dvar) \\[1mm]
\vec{F}\in \vars^* \mbox{ vector of variables} & P \in \Pp \mbox{ predicates} \\[1mm]
x\in\vars\mbox{ variable} \hspace{1cm}& \Delta\ \mbox{ formula over data variables} 
\end{array}
\\[0.7cm]
\Pi\ ::=\ X = Y \mid X \neq Y \mid \Delta\mid \Pi \land \Pi \hfill \mbox{pure formulas}\\[0.7mm]
\Sigma\ ::=\ 
\slemp \mid 
E \mapsto \rho \mid P(E,\vec{F}) \mid \Sigma \sep \Sigma 
\hspace{1cm}\hfill \mbox{spatial formulas}
\\[1.2mm]
\varphi\ ::=\ \Pi \land \Sigma\mid \varphi\vee \varphi\mid \exists x.\ \varphi \hfill \mbox{formulas} %
\end{array}
$$
\vspace{-6eX}
\end{table}

Formulas are conjunctions between a pure formula $\Pi$ and a spatial formula $\Sigma$.
Pure formulas characterize the stack $s$ using (dis)equalities between location variables, e.g., a stack models $x=y$ iff $s(x)=s(y)$,
and constraints $\Delta$ over data variables.
We let $\Delta$ unspecified, though we assume that they belong to decidable
theories, e.g., linear arithmetic or quantifier-free first order theories over multisets of values. 
The atom $\slemp$ of spatial formulas holds iff the domain of the heap is empty.
The \emph{points-to atom} $E \mapsto \{ (f_i,x_i) \}_{i\in\mathcal{I}}$
specifies that the heap contains exactly one location $E$, and for all $i\in \mathcal{I}$, the field $f_i$ of $E$ equals $x_i$, i.e., $h(s(E),f_i)=s(x_i)$.
The \emph{predicate atom} $P(E,\vec{F})$ specifies a heap segment rooted at $E$ and shaped by the predicate $P$;
the fragment is parameterized by a set $\Pp$
of \emph{inductively defined predicates}, formally defined hereafter.

Let $P \in \Pp$. An \emph{inductive definition} of $P$ is a finite set of rules of the form 
$P(E,\vec{F})::=\exists \vec{Z}. \Pi \land \Sigma$, where $\vec{Z} \in \vars^\ast$ is a tuple of variables.
A rule $R$ is called a \emph{base rule} if $\Sigma$ contains no predicate atoms. Otherwise, it is called an \emph{inductive rule}.
%
A base rule $R$ is called \emph{spatial-empty} if $\Sigma=\slemp$. Otherwise, it is called a \emph{spatial-nonempty} base rule. For instance, the predicate $\pbst$ in Sec.~\ref{sec:motivation} is defined by one spatial-empty base rule and one inductive rule. 

We consider a class of restricted inductive definitions that are expressive enough to deal
with intricate data structures (see Sec.~\ref{sec:exp}) while also enabling efficient proof
strategies 
for establishing the validity of the verification conditions (see Sec.~\ref{sec:slice}). 
%
For each rule $R: P(E,\vec{F})::=\exists \vec{Z}. \Pi \land \Sigma$ in the definition of a predicate $P(E,\vec{F}) \in \Pp$, we assume that:
\vspace{-2mm}
\begin{itemize}
\item If $R$ is inductive, then $\Sigma = \Sigma_1 \sep \Sigma_2$ and the following conditions hold:
\begin{itemize}
\item \emph{the root atoms}: $\Sigma_1$ contains only points-to atoms and 
a \emph{unique} points-to atom starting from $E$, denoted as $E \mapsto \rho$. Also, all the \emph{location} variables from $\vec{Z}$ occur in $\Sigma_1$.
$\Sigma_1$ is called the \emph{root} of $R$ and denoted by $root(R)$.
\item \emph{connectedness}: the Gaifman graph of $\Sigma_1$, denoted by $G_{\Sigma_1}$, is 
a connected DAG (directed acyclic graph) with the root $E$, 
that is, every vertex is reachable from $E$, 
%
%
%
\item \emph{predicate atoms}: $\Sigma_2$ contains only atoms of the form $Q(Z,\vec{Z'})$, and for each such atom, $Z$ is 
a vertex in $G_{\Sigma_1}$ without outgoing arcs.
%
%
%
\end{itemize}
%
\item If $R$ is a spatial-nonempty base rule, then $\Sigma$ contains exactly one points-to atom $E \mapsto \rho$, for some $\rho$.
\end{itemize}

\vspace{-1mm}
\noindent
The classic acyclic list segment definition~\cite{Rey02} satisfies these constraints as well as 
the first rule below; the second rule below falsifies the ``root atoms'' constraint:
\vspace{-1eX}
{\small
\begin{align*}
\plsegeven(E,F) &::=\exists X,Y.\ E \mapsto (\fnext, X) \sep X \mapsto (\fnext,Y) \sep \plsegeven(Y,F)
\\[1mm]
\plsegb(E,F) &::=\exists X.\ \plsegb(E,X) \sep X \mapsto (\fnext,F).
\end{align*}}

\vspace{-6mm}
\noindent
Since we disallow the use of negations on top of the spatial atoms, the semantics of the predicates in $\Pp$ is defined as usual as a least fixed-point. 
The class of inductive definitions defined above is in general undecidable, since with data fields, inductive definitions can be used to simulate two-counter machines. 

A \emph{variable substitution} $\eta$ is a mapping from a finite subset of $\vars$ to the set of terms over the respective domains. 
For instance, if $X \in \lvar$ and $v,v_1 \in \dvar$ be integer variables then  
the mapping $\eta=\{X \rightarrow \nil, v \rightarrow v_1 + 5\}$ is a variable substitution. 
We denote by $\freev(\psi)$  the set of free variables of a formula $\psi$. 

\vspace{-3eX}

\section{Composition Lemmas}\label{sec:comp}

\vspace{-2eX}

As we have seen in the motivating example, the predicate $\pbsthole(E,M_1,F,M_2)$ satisfies the property that composing two heap structures described by this predicate results in a heap structure satisfying the same predicate. We call this property a \emph{composition lemma}. We define simple and uniform syntactic criteria which, if they are satisfied by a predicate, then the composition lemma holds. 

The main idea is to divide the parameters of inductively defined predicates into three categories: The \emph{source} parameters $\orar{\alpha}=(E,C)$, the \emph{hole} parameters $\orar{\beta}=(F,H)$, and the \emph{static} parameters $\orar{\xi}\in \vars^\ast$, where $E,F \in \lvar$ are called the source and resp., the hole location parameter, and $C,H \in \dvar$ are called the cumulative and resp., the hole data parameter\footnote{For simplicity, we assume that  $\orar{\alpha}$ and $\orar{\beta}$ consist of exactly one location parameter and one data parameter.}.


%
%
%
%
%
%

Let $\Pp$ be a set of inductively defined predicates and $P \in \Pp$ with the parameters $(\orar{\alpha},\orar{\beta},\orar{\xi}\,)$. Then $P$ is said to be \emph{syntactically compositional} if 
the inductive definition of $P$ contains \emph{exactly one base rule}, and \emph{at least one inductive rule}, and the rules of $P$ are of one of the following forms:
\vspace{-2mm}
\begin{itemize}
\item Base rule: $P(\orar{\alpha},\orar{\beta},\orar{\xi}\,)::= \alpha_{1} = \beta_{1} \land \alpha_2 = \beta_2 \land \slemp$. Note that here the points-to atoms are disallowed.
\item Inductive rule:  
$P(\orar{\alpha},\orar{\beta},\orar{\xi}\,)::= \exists \orar{Z}. \ \Pi \land \Sigma$, with 
(a) $\Sigma \triangleq \Sigma_1 \sep \Sigma_2 \sep P(\orar{\gamma},\orar{\beta},\orar{\xi}\,)$, 
(b) $\Sigma_1$ contains only and at least one points-to atoms,
(c) $\Sigma_2$ contains only and possibly none predicate atoms,
(d) $\vec{\gamma} \subseteq \orar{Z}$, and
(d) the variables in $\orar{\beta}$ \emph{do not occur elsewhere} in $\Pi \land \Sigma$, i.e., not in $\Pi$, or $\Sigma_1$, or $\Sigma_2$, or $\vec{\gamma}$. Note that the inductive rule also satisfies the constraints ``root atom'' and ``connectedness'' introduced in Sec.~\ref{sec:logic}. In addition, $\Sigma_2$ may contain $P$ atoms.
\end{itemize}
\vspace{-2mm}


One may easily check that both the predicate $\plseg(E,M,F,M')$ in eq.~(\ref{rule:lseg-base})--(\ref{rule:lseg-rec}) and  
the predicate $\pbsthole(E,M_1,F,M_2)$ in eq.~(\ref{rule:bsth-base})--(\ref{rule:bsth-recl}) are syntactically compositional, 
while the predicate $\plseg(E,M,F)$ in eq.~(\ref{rule:lseg1-base})--(\ref{rule:lseg1-rec}) is not.

A predicate $P \in \Pp$ with the parameters $(\orar{\alpha},\orar{\beta},\orar{\xi}\,)$ 
is said to be \emph{semantically compositional} if the entailment $\exists \orar{\beta}.\ P(\orar{\alpha},\orar{\beta},\orar{\xi}\,) \sep P(\orar{\beta},\orar{\gamma},\orar{\xi}\,) \Rightarrow P(\orar{\alpha}, \orar{\gamma},\orar{\xi}\,)$ holds.

\vspace{-2mm}
\begin{theorem}\label{thm-one-pred-compos}
Let $\Pp$ be a set of inductively defined predicates. If $P \in \Pp$ is syntactically compositional, then $P$ is semantically compositional. 
\end{theorem}
\vspace{-2mm}
The proof of Thm.~\ref{thm-one-pred-compos} is done (see~\cite{ESZ15}) by induction on the size of the domain of the heap structures. Suppose $(s,h) \models P(\orar{\alpha},\orar{\beta},\orar{\xi}\,) \sep P(\orar{\beta},\orar{\gamma},\orar{\xi}\,)$, then either $s(\orar{\alpha})=s(\orar{\beta})$ or $s(\orar{\alpha}) \neq s(\orar{\beta})$. If the former situation occurs, then $(s,h) \models P(\orar{\alpha}, \orar{\gamma},\orar{\xi}\,)$ follows immediately. Otherwise, 
the predicate $P(\orar{\alpha},\orar{\beta},\orar{\xi}\,)$ is unfolded by using some inductive rule of $P$, and the induction hypothesis can be applied to a sub-heap of smaller size. Then $(s,h) \models P(\orar{\alpha}, \orar{\gamma},\orar{\xi}\,)$ can be deduced by utilizing the property that the hole parameters occur only once in each inductive rule of $P$.


\vspace{-1eX}
\begin{remark}
The syntactically compositional predicates are rather general in the sense that they allow nestings of predicates, branchings (e.g. trees), as well as data and size constraints. Therefore, composition lemmas can be obtained for complex data structures like nested lists, AVL trees, red-black trees, and so on. In addition, although lemmas have been widely used in the literature, we are not aware of any work that uses the composition lemmas as simple and elegant as those introduced above, when data and size constraints are included.
\end{remark}


\vspace{-7mm}

\section{Derived Lemmas}\label{sec:derived}

\vspace{-3mm}

%

Theorem~\ref{thm-one-pred-compos} provides a mean to obtain lemmas for one single syntactically compositional predicate. In the following, based on the syntactic compositionality, we demonstrate how to derive additional lemmas describing relationships between different predicates (proofs are detailed in~\cite{ESZ15}). 
We identify three categories of derived lemmas: ``completion'' lemmas, ``stronger'' lemmas, and ``static-parameter contraction'' lemmas. 
Based on our experiences in the experiments (cf. Sec.~\ref{sec:exp}) and the examples from the literature, 
we believe that the composition lemmas as well as the derived ones are natural, essential, and general enough for the verification of programs manipulating dynamic data structures. 
For instance, the ``composition'' lemmas and ``completion'' lemmas are widely used in our experiments, the ``stronger'' lemmas are used to check the verification conditions for rebalancing AVL trees and red-black trees. While ``static parameter contraction'' lemmas are not used in our experiments, they could also be useful, e.g., for the verification of programs manipulating lists with tail pointers. 

\vspace{-4mm}
\subsection{The ``completion'' lemmas}
\vspace{-1mm}

We first consider the ``completion'' lemmas which describe relationships between incomplete data structures (e.g., binary search trees with one hole) and complete data structures (e.g., binary search trees). 
For example, the following lemma is valid for the predicates $\pbsthole$ and $\pbst$: 
\vspace{-1mm}
{\small
\begin{align*}
&\exists F,M_2.\ \pbsthole(E,M_1, F, M_2) \sep \pbst(F,M_2) \Rightarrow \pbst(E,M_1).
\end{align*}}

\vspace{-6mm}
\noindent
Notice that the rules defining $\pbst(E,M)$ can be obtained from those of $\pbsthole(E_1,M_1,F,M_2)$ by applying the variable substitution $\eta=\{F \rightarrow \nil, M_2 \rightarrow \emptyset\}$ (modulo the variable renaming $M_1$ by $M$). This observation is essential to establish the ``completion lemma'' and it is generalized to arbitrary syntactically compositional predicates as follows.

Let $P \in \Pp$ be a syntactically compositional predicate with the parameters $(\orar{\alpha},\orar{\beta},\orar{\xi}\,)$, and $P' \in \Pp$ a predicate with the parameters $(\orar{\alpha},\orar{\xi}\,)$. 
Then $P'$ is a \emph{completion} of $P$ with respect to a pair of constants $\orar{c}=c_1c_2$, if the rules of $P'$ are obtained from the rules of $P$ by applying the variable substitution $\eta=\{\beta_1 \rightarrow c_1, \beta_2 \rightarrow c_2\}$. More precisely, 
\vspace{-1mm}
\begin{itemize}
\item let $\alpha_1 = \beta_1 \land \alpha_2 = \beta_2 \land \slemp$ be the base rule of $P$, then $P'$ contains only one base rule, that is, $\alpha_1  = c_1 \land \alpha_2 = c_2 \land \slemp$,
\item the set of inductive rules of $P'$ is obtained from those of $P$ as follows: Let $P(\orar{\alpha},\orar{\beta},\orar{\xi}\,)::=\exists \orar{Z}.\ \Pi \land \Sigma_1 \sep \Sigma_2 \sep P(\orar{\gamma},\orar{\beta},\orar{\xi}\,)$ be an inductive rule of $P$, then $P'(\orar{\alpha},\orar{\xi}\,)::=\exists \orar{Z}.\ \Pi \land \Sigma_1 \sep \Sigma_2 \sep P'(\orar{\gamma},\orar{\xi}\,)$ is an inductive rule of $P'$ (Recall that $\vec{\beta}$ does not occur in $\Pi,\Sigma_1,\Sigma_2,\orar{\gamma}$).
%
%
\end{itemize}
\vspace{-1mm}


\vspace{-1mm}
\begin{theorem}\label{thm-completion}
Let $P(\orar{\alpha},\orar{\beta},\orar{\xi}\,) \in \Pp$ be a syntactically compositional predicate, and $P'(\orar{\alpha},\orar{\xi}\,) \in \Pp$. If $P'$ is a completion of $P$ with respect to $\orar{c}$, then $P'(\orar{\alpha},\orar{\xi}\,) \Leftrightarrow P(\orar{\alpha},\orar{c},\orar{\xi}\,)$ and  $\exists\orar{\beta}.\ P(\orar{\alpha},\orar{\beta},\orar{\xi}\,) \sep P'(\orar{\beta},\orar{\xi}\,) \Rightarrow P'(\orar{\alpha}, \orar{\xi}\,)$ hold.
\end{theorem}


\vspace{-7mm}
\subsection{The ``stronger'' lemmas}
\vspace{-1mm}

We illustrate this class of lemmas on the example of binary search trees. 
Let ${\it natbsth}(E,M_1, F, M_2)$ be the predicate defined by the same rules as $\pbsthole(E,M_1,F,M_2)$  (i.e., eq.~(\ref{rule:bsth-base})--(\ref{rule:bsth-recl})), 
except that $M_3 \ge 0$ ($M_3$ is an existential variable) is added to the body of each inductive rule (i.e., eq.~(\ref{rule:bsth-recr}) and (\ref{rule:bsth-recl})). 
Then we say that ${\it natbsth}$ is \emph{stronger} than $\pbsthole$, since for each rule $R'$ of ${\it natbsth}$, there is a rule $R$ of $\pbsthole$, such that the body of $R'$ entails the body of $R$. This ``stronger'' relation guarantees that the following lemmas hold:
\vspace{-1mm}
{\small
\begin{align*}
{\it natbsth}(E,M_1,F,M_2) & \limp \pbsthole(E,M_1,F,M_2) 
\\
\exists E_2,M_2.\ {\it natbsth}(E_1, M_1, E_2, M_2) \sep \pbsthole(E_2, M_2, E_3, M_3) & \limp 
\pbsthole(E_1,M_1, E_3, M_3).
\end{align*}
}

\vspace{-3eX}
\noindent
In general, for two syntactically compositional predicates $P,P' \in \Pp$ with the same set of parameters $(\orar{\alpha},\orar{\beta},\orar{\xi}\,)$, $P'$ is said to be \emph{stronger} than $P$ if 
for each inductive rule 
$P'(\orar{\alpha},\orar{\beta},\orar{\xi}\,)::=\exists \orar{Z}. \ \Pi' \land \Sigma_1 \sep \Sigma_2 \sep P'(\orar{\gamma},\orar{\beta},\orar{\xi}\,)$, 
there is an inductive rule $P(\orar{\alpha},\orar{\beta},\orar{\xi}\,)::=\exists \orar{Z}. \ \Pi \land \Sigma_1 \sep \Sigma_2 \sep P(\orar{\gamma},\orar{\beta},\orar{\xi}\,)$ such that 
$\Pi' \limp \Pi$ holds.
The following result is a consequence of Thm.~\ref{thm-one-pred-compos}.
\vspace{-1mm}
\begin{theorem}\label{thm-strong}
Let $P(\orar{\alpha},\orar{\beta},\orar{\xi}\,),P'(\orar{\alpha},\orar{\beta},\orar{\xi}\,) \in \Pp$ be two syntactically compositional predicates. If $P'$ is stronger than $P$, then the entailments 
$P'(\orar{\alpha},\orar{\beta},\orar{\xi}\,)\limp P(\orar{\alpha},\orar{\beta},\orar{\xi}\,)$ 
and 
$\exists\orar{\beta}.\ P'(\orar{\alpha},\orar{\beta},\orar{\xi}\,) \sep P(\orar{\beta},\orar{\gamma},\orar{\xi}\,) \limp P(\orar{\alpha},\orar{\gamma},\orar{\xi}\,)$ hold.
\end{theorem}
\vspace{-1mm}
The ``stronger'' relation defined above requires that the spatial formulas in the inductive rules of $P$ and $P'$ are the same. 
This constraint can be relaxed by only requiring that the body of each inductive rule of $P'$ is stronger than a formula obtained by unfolding an inductive rule of $P$ for a \emph{bounded number of times}. 
This relaxed constraint allows generating additional lemmas, e.g., the lemmas relating the predicates for list segments of even length and list segments.


\vspace{-4mm}
\subsection{The ``static-parameter contraction'' lemmas}
\vspace{-1mm}



Let ${\it tailbsth}(E,M_1,F,M_2)$ (resp. ${\it stabsth}(E,M_1,F, M_2,B)$) be the predicate defined by the same rules as $\pbsthole(E,M_1,F,M_2)$, with the modification that the points-to atom in each inductive rule is replaced by $E \mapsto \{(\fleft,X),(\fright,Y),({\tt tail},F),(\fdata,v)\}$ (resp. $E \mapsto \{(\fleft,X),(\fright,Y),({\tt tail},B),(\fdata,v)\}$). Intuitively, ${\it tailbsth}$ (resp. ${\it stabsth}$) is obtained from $\pbsthole$ by adding a ${\tt tail}$ pointer to $F$ (resp. $B$).
Then ${\it tailbsth}$ is not syntactically compositional since $F$ occurs in the points-to atoms of the inductive rules. On the other hand, ${\it stabsth}$ is syntactically compositional.

From the above description, it is easy to observe that the inductive definition of ${\it tailbsth}(E, M_1, F, M_2)$ can be obtained from that of ${\it stabsth}(E, M_1, F, M_2, B)$ by replacing $B$ with $F$. 
Then the lemma ${\it tailbsth}(E, M_1, F, M_2) \Leftrightarrow {\it stabsth}(E, M_1, F, M_2, F)$ holds. From this, we further deduce the lemma
\vspace{-1mm}
{\small
\begin{align*}
& \exists E_2, M_2.\ {\it stabsth}(E_1, M_1, E_2, M_2, E_3) \sep {\it tailbsth}(E_2, M_2, E_3, M_3) \limp \\
&\hspace{7.7cm} 
{\it tailbsth}(E_1, M_1, E_3, M_3).
\end{align*}
}

\vspace{-6mm}
\noindent
We call the aforementioned replacement of $B$ by $F$ in the inductive definition of ${\it stabsth}$ as the ``static-parameter contraction''. This idea can be generalized to arbitrary syntactically compositional predicates as follows.


Let $P \in \Pp$ be a syntactically compositional predicate with the parameters $(\orar{\alpha}, \orar{\beta}, \orar{\xi}\,)$, $P' \in \Pp$ be an inductive predicate with the parameters $(\orar{\alpha}, \orar{\beta}, \orar{\xi'}\,)$, $\orar{\xi}=\xi_1\dots \xi_k$, and $\orar{\xi'}=\xi'_1 \dots \xi'_l$. Then $P'$ is called a \emph{static-parameter contraction} of $P$ if the rules of $P'$ are obtained from those of $P$ by a variable substitution $\eta$ s.t. $\dom(\eta)=\vec{\xi}$, for each $i: 1 \le i \le k$, either $\eta(\xi_i)=\xi_i$, or $\eta(\xi_i)=\beta_j$ for some $j=1,2$ satisfying that $\xi_i$ and $\beta_j$ have the same data type, and $\orar{\xi'}$ is the tuple obtained from $\eta(\vec{\xi}~)$ by removing the $\beta_j$'s. The substitution $\eta$ is called the \emph{contraction function}.

\vspace{-1mm}
\begin{theorem}\label{thm-border-contr}
Let $P(\orar{\alpha}, \orar{\beta}, \orar{\xi}\,) \in \Pp$ be a syntactically compositional predicate and $P'(\orar{\alpha}, \orar{\beta}, \orar{\xi'}) \in \Pp$ be an inductive predicate. If $P'$ is a static-parameter contraction of $P$ with the contraction function $\eta$, then $P'(\orar{\alpha}, \orar{\beta}, \orar{\xi'}) \Leftrightarrow P(\orar{\alpha}, \orar{\beta}, \eta(\vec{\xi}\ ))$ and $\exists\orar{\beta}.\ P(\orar{\alpha}, \orar{\beta}, \eta(\vec{\xi}\ )) \sep P'(\orar{\beta}, \orar{\gamma}, \orar{\xi'}) \Rightarrow P'(\orar{\alpha}, \orar{\gamma}, \orar{\xi'})$ hold.
\end{theorem}
\vspace{-3mm}

\begin{remark}
The lemmas presented in the last two sections are incomplete in the sense that they may not cover all the lemmas for a given set of inductive predicates. Although various extensions of the lemmas are possible, generating all the possible lemmas can be quite complex in general. Thm. 3 in \cite{IRV14} shows that generating all the lemmas is at least EXPTIME-hard, even for a fragment restricted to shape properties, without any data or size constraint.
\end{remark}



\vspace{-5mm}

\section{A Proof Strategy Based on Lemmas}
\label{sec:slice}
\vspace{-1mm}

%
%
%
%

%
%
%

We introduce a proof strategy based on lemmas for proving entailments $\varphi_1\limp\exists\vec{X}.\varphi_2$, where $\varphi_1$, $\varphi_2$ are quantifier-free, and $\vec{X} \in \dvar^\ast$. 
The proof strategy treats uniformly the inductive rules defining predicates and the lemmas defined in Sec.~\ref{sec:comp}--\ref{sec:derived}. 
%
Therefore, we call lemma also an inductive rule.
W.l.o.g. we assume that $\varphi_1$ is quantifier-free (the existential variables can be skolemized). 
In addition, we assume that \emph{only data variables are quantified in the right-hand side}\footnote{We believe that this restriction is reasonable for the verification conditions appearing in practice and all the benchmarks in our experiments are of this form.}.

W.l.o.g., we assume that every variable in $\vec{X}$ occurs in at most one spatial atom of $\varphi_2$ (multiple occurrences of the same variable can be removed by introducing fresh variables and new equalities in the pure part). 
Also, we assume that $\varphi_1$ and $\varphi_2$ are of the form $\Pi\land\Sigma$. 
In the general case, our proof strategy checks that for every disjunct $\varphi_1'$ of $\varphi_1$, there is a disjunct $\varphi_2'$ of $\varphi_2$ s.t. $\varphi_1'\limp\exists \vec{X}.\varphi_2'$.

We present the proof strategy as a set of rules in Fig.~\ref{fig-pf-rule}.
For a variable substitution $\eta$ and a set $\mathcal{X} \subseteq \vars$, we denote by $\eta|_{\mathcal{X}}$ the restriction of $\eta$ to $\mathcal{X}$. 
In addition, $\eqmap(\eta)$ is the conjunction of the equalities $X=t$ for every $X$ and $t$ such that $\eta(X)=t$. 
Given two formulas $\varphi_1$ and $\varphi_2$, a substitution $\eta$ with $\dom(\eta)=\vec{X}$, 
the judgement 
$\varphi_1\models_{\eta} \exists\vec{X}.\varphi_2$ denotes that the entailment 
$\varphi_1 \limp \eta(\varphi_2)$ is valid. 
Therefore, $\eta$ provides an instantiation for the quantified variables $\vec{X}$ which witnesses the validity.

\begin{figure}[t]
\vspace{-3mm}
\fbox{
\begin{minipage}[t]{0.98\textwidth}
\fbox{
\begin{minipage}[h]{0.96\textwidth}
\footnotesize
\[(\textsc{Match1})\hspace{2mm} \frac{\Sigma_1 = \theta(\Sigma_2) \hspace{6mm} \eta=\theta|_{\vec{X}}\hspace{6mm} \Pi_1 \land \eqmap(\eta) \models \eqmap(\theta|_{\freev(\exists \vec{X}.\Sigma_2)}) } {\Pi_1 \land \Sigma_1 \models_{\eta}^{\rSUB} \exists \vec{X}.\  \Sigma_2}
\]
\end{minipage}
}
\begin{minipage}[h]{0.96\textwidth}
\ 
\end{minipage}
\fbox{
\begin{minipage}[h]{0.96\textwidth}
\footnotesize
\[(\textsc{Match2})\hspace{4mm} \frac{\Pi_1 \land \Sigma_1 \models_{\eta}^{\rSUB} \exists \vec{X}.\  \Sigma_2} {\Pi_1 \land \Sigma_1 \models_{\eta} \exists \vec{X}.\  \Sigma_2}
\]
\end{minipage}
}
\begin{minipage}[h]{0.96\textwidth}
\ 
\end{minipage}
\fbox{
\begin{minipage}[h]{0.96\textwidth}
\footnotesize
\[
(\textsc{Lemma})\hspace{2mm} \frac{
\Pi_1 \land \Sigma_1 \models^{\rSUB}_{\eta_1} \exists\vec{Z'}.\ root(L) 
\hspace{3mm} \Pi_1 \land \Sigma_1' \models_{\eta_2} \exists \vec{Z''}.\ \eta_1(\Pi \land \Sigma)
}{\Pi_1 \land \Sigma_1 \sep \Sigma_1' \models_{\eta|_{\vec{X}}} \exists \vec{X}.\ A}
\]
%
\begin{itemize}
\item $L ::= \exists \vec{Z}.\ \Pi \land root(L) \sep \Sigma \Rightarrow A$ is a lemma,
\item $\vec{Z'}=(\vec{X} \cup \vec{Z}) \cap \freev(root(L))$, $\vec{Z''}=(\vec{X} \cup \vec{Z}) \cap \freev(\eta_1(\Pi \land \Sigma))$, 
\item $\eta=\ext_{\Pi}(\eta_1 \cup \eta_2)$ is the extension of $\eta_1 \cup \eta_2$ with $\Pi$ s.t. $\dom(\eta)=\vec{X} \cup \vec{Z}$.
\end{itemize}
\end{minipage}
}
\begin{minipage}[h]{0.96\textwidth}
\ 
\end{minipage}
\fbox{
\begin{minipage}[h]{0.96\textwidth}
\footnotesize
\[(\textsc{Slice})\hspace{2mm} \frac{\Pi_1 \land \Sigma_1 \models_{\eta_1} \exists \vec{Z'}. A \hspace{3mm} \Pi_1 \land \Sigma_2 \models_{\eta_2} \exists \vec{Z''}. \Sigma \hspace{3mm} \Pi_1 \land  \eqmap(\eta) \models \Pi_2 } {\Pi_1 \land \Sigma_1 \sep \Sigma_2 \models_{\eta} \exists \vec{X}.\ \Pi_2 \land A \sep \Sigma}\]
\begin{itemize}
\item $\vec{Z'}=\vec{X} \cap \freev(A)$, $\vec{Z''}=\vec{X} \cap \freev(\Sigma)$, 
\item $\eta=\ext_{\Pi_2}(\eta_1 \cup \eta_2)$ is the extension of $\eta_1 \cup \eta_2$ with $\Pi_2$ s.t. $\dom(\eta)=\vec{X}$.
\end{itemize}
\end{minipage}
}
\end{minipage}
}
\vspace{-2mm}
\caption{The proof rules for checking the entailment $\varphi_1 \Rightarrow \exists \vec{X}.\ \varphi_2$}
\label{fig-pf-rule}
\vspace{-4eX}
\end{figure}


The rules $\textsc{Match1}$ and $\textsc{Match2}$ consider a particular case of $\models_{\eta}$, denoted using the superscript $\rSUB$, where the spatial atoms of $\varphi_2$ are syntactically matched\footnote{In this case, the right-hand side contains no pure constraints.} 
to the spatial atoms of $\varphi_1$ modulo a variable substitution $\theta$. 
The substitution of the existential variables is recorded in $\eta$, 
while the substitution of the free variables generates a set of equalities that must be implied by $\Pi_1\land \eqmap(\eta)$.
For example, let
%
$\Pi_1 \land \Sigma_1  ::= w=w'\land E\mapsto\{(f,Y),(d_1,v),(d_2,w)\}$, and  
$\exists \vec{X}.\ \Sigma_2   ::= \exists X,v'.\ E\mapsto\{(f,X),(d_1,v'),(d_2,w')\}$,
%
where $d_1$ and $d_2$ are data fields. If $\theta=\{X \rightarrow Y, v' \rightarrow v, w' \rightarrow w\}$, then $\Sigma_1 = \theta(\Sigma_2)$. The substitution of the free variable $w'$ from the right-hand side is sound since the equality $w=w'$ occurs in the left-hand side.
Therefore, $\Pi_1 \land \Sigma_1 \models^{SUB}_{\theta|_{\{X,v'\}}} \exists X, v'.\ \Sigma_2$ holds. 


The rule $\textsc{Lemma}$ applies a lemma $L::= \exists \vec{Z}.\ \Pi \land root(L) \sep \Sigma \limp A$.
It consists in proving that $\varphi_1$ implies the LHS of the lemma where the variables in $\vec{X}$ are existentially quantified, i.e., $\exists\vec{X}\exists \vec{Z}.\ \Pi \land root(L) \sep \Sigma$.
Notice that $\vec{Z}$ may contain existential location variables. 
Finding suitable instantiations for these variables relies on the assumption
that ${\it root}(L)$ in the LHS of $L$ is either a \emph{unique predicate atom} or a \emph{separating conjunction of points-to atoms} rooted at $E$ (the first parameter of $A$)  
and ${\it root}(L)$ includes all the location variables in $\vec{Z}$. 
This assumption holds for all the inductive rules defining predicates in our fragment (a consequence of the root and connectedness constraints) and for all the lemmas defined in Sec.~\ref{sec:comp}--\ref{sec:derived}.
The proof that $\varphi_1$ implies $\exists\vec{X}\exists \vec{Z}.\ \Pi \land root(L) \sep \Sigma$ is split into two sub-goals 
(i) proving that a sub-formula of $\varphi_1$ implies $\exists\vec{X}\exists \vec{Z}.\ root(L)$ and 
(ii) proving that a sub-formula of $\varphi_1$ implies $\exists\vec{X}\exists \vec{Z}.\ \Pi\land \Sigma$. 
The sub-goal (i) relies on syntactic matching using the rule $\textsc{Match1}$, which results in a quantifier instantiation $\eta_1$.
The substitution $\eta_1$ is used to instantiate existential variables in $\exists \vec{X} \exists \vec{Z}.\ \Pi \land \Sigma$. 
Notice that according to the aforementioned assumption, the location variables in $\vec{Z}$ are not free in $\eta_1(\Pi \land \Sigma)$.
Let $\eta_2$ be the quantifier instantiation obtained from the second sub-goal.
The quantifier instantiation $\eta$ is defined as the extension of $\eta_1 \cup \eta_2$ to the domain $\vec{X} \cup \vec{Z}$ by utilizing the pure constraints $\Pi$ from the lemma\footnote{The extension depends on the pure constraints $\Pi$ and could be quite complex in general. In the experiments of Sec.~\ref{sec:exp}, we use the extension obtained by the propagation of equalities in $\Pi$.}. 
This extension is necessary since some existentially quantified variables may only occur in $\Pi$, but not in $root(L)$ nor in $\Sigma$, so  they are not covered by $\eta_1 \cup \eta_2$. For instance, if $\Pi$ contains a conjunct $M=M_1 \cup M_2$ such that $M_1\in \dom(\eta_1)$, $M_2  \in \dom(\eta_2)$, and $M \not \in \dom(\eta_1 \cup \eta_2)$, then $\eta_1 \cup \eta_2$ is extended to $\eta$ where $\eta(M)=\eta_1(M_1) \cup \eta_2(M_2)$. 

The rule $\textsc{Slice}$ chooses a spatial atom $A$ in the RHS and generates two sub-goals:
(i) one that matches $A$ (using the rules $\textsc{Match2}$ and $\textsc{Lemma}$)
with a spatial sub-formula of the LHS ($\Sigma_1$) and
(ii) another that checks that the remaining spatial part of the RHS is implied by the remaining part of the LHS.
%
The quantifier instantiations $\eta_1$ and $\eta_2$ obtained from the two sub-goals are used to check that the pure constraints in the RHS are implied by the ones in LHS. Note that in the rule $\textsc{Slice}$, it is possible that $\Sigma_2=\Sigma=\slemp$.

The rules in Fig.~\ref{fig-pf-rule} are applied in the order given in the figure. Note that they focus on disjoint cases w.r.t. the syntax of the RHS.
The choice of the atom $A$ in $\textsc{Slice}$ is done arbitrary, since it does not affect the efficiency of proving validity. 

We apply the above proof strategy to the entailment 
$\varphi_1\limp\exists M.\ \varphi_2$ where:
\vspace{-1eX}
{\small
\begin{align*}
\varphi_1 & ::= x_1\neq\nil\land x_2\neq \nil \land v_1 < v_2\land x_1 \mapsto \{(\fnext,x_2),(\fdata,v_1)\} \\
& \hspace{34eX}\sep\ x_2 \mapsto \{(\fnext,\nil), (\fdata,v_2)\} \\
\varphi_2 & ::= \plseg(x_1, M, \nil,\emptyset) \land v_2\in M,
\end{align*}
}

\vspace{-5mm}
\noindent
and $\plseg$ has been defined in Sec.~\ref{sec:intro} (eq.~(\ref{rule:lseg-base})--(\ref{rule:lseg-rec})).  
The entailment is valid because it states that two cells linked by $\fnext$ and storing ordered data values form a sorted list segment.
The RHS $\varphi_2$ contains a single spatial atom and a pure part so the rule $\textsc{Slice}$ is applied and it generates the sub-goal $\varphi_1\models_{\eta}\exists M.\ \plseg(x_1, M, \nil,\emptyset)$ for which the syntactic matching (rule \textsc{Match1}) can not be applied. 
Instead, we apply the rule $\textsc{Lemma}$ using as lemma the inductive rule of $\plseg$,
%
%
%
i.e., eq.~(\ref{rule:lseg-rec}) (page~\pageref{rule:lseg-rec}).
We obtain the RHS 
$\exists M,X,M_1,v.\ x_1 \mapsto \{(\fnext,X), (\fdata,v)\} \sep \plseg(X,M_1,\nil,\emptyset) 
\land M=\{v\}\cup M_1 \land v \le M_1$%
, where
$x_1 \mapsto \{(\fnext,X), (\fdata,v)\}$ is the root.
The rule $\textsc{Match1}$ is applied 
with 
$\Pi_1 \land \Sigma_1 ::= x_1\neq\nil\land x_2\neq \nil \land v_1 < v_2\land x_1 \mapsto \{(\fnext,x_2),(\fdata,v_1)\}$ 
%
and it returns 
the substitution $\eta_1=\{X \rightarrow x_2, v \rightarrow v_1\}$.
The second sub-goal is $\Pi_1 \land \Sigma_2 \models_{\eta_2} \exists M,M_1. \psi' $ where
$\Pi_1 \land \Sigma_2 ::= x_1\neq\nil \land x_2\neq \nil \land v_1 < v_2 \land x_2 \mapsto \{(\fnext,\nil), (\fdata,v_2)\}$
and $\psi' ::= M=\{v_1\} \cup M_1 \land v_1 \le M_1 \land \plseg(x_2,M_1,\nil,\emptyset)$. 
For this sub-goal, we apply the rule $\textsc{Slice}$, which generates a sub-goal where the rule $\textsc{Lemma}$ is applied first, using the same lemma, then the rule \textsc{Slice} is applied again, and finally the rule \textsc{Lemma} is applied with a lemma corresponding to the base rule of $\plseg$, i.e., eq.~(\ref{rule:lseg-base}) (page~\pageref{rule:lseg-rec}). 
This generates a quantifier instantiation $\eta_2=\{M \rightarrow \{v_1,v_2\}, M_1 \rightarrow \{v_2\}\}$. 
Then, $\eta_1 \cup \eta_2$ is extended with the constraints from the pure part of the lemma, i.e., 
$M=\{v\}\cup M_1 \land v_1 \le M_1$. Since $M \in \dom(\eta_1 \cup \eta_2)$, this extension has no effect.  
%
%
Finally, the rule $\textsc{Slice}$ checks that 
$\Pi_1 \land \eqmap(\eta|_{\{M\}}) \models \Pi_2$ holds, where 
$\eqmap(\eta|_{\{M\}}) ::= M = \{v_1,v_2\}$ and $\Pi_2 ::= v_2 \in M$. 
The last entailment holds, so the proof of validity is done.
The following theorem states the correctness of the proof rules. Moreover, since we assume a finite set of lemmas, and every application of a lemma $L$ removes at least one spatial atom from $\varphi_1$ (the atoms matched to ${\it root}(L)$), the termination of the applications of the rule $\textsc{Lemma}$ is guaranteed. 

\vspace{-2mm}
\begin{theorem}
Let $\varphi_1$ and $\exists\vec{X}.\varphi_2$ be two formulas such that $\vec{X}$ contains only data variables. 
If $\varphi_1 \models_{\eta} \exists\vec{X}.\varphi_2$ for some $\eta$,  then $\varphi_1\limp \exists\vec{X}.\varphi_2$.
\end{theorem}


\vspace{-5eX}
\section{Experimental results}
\label{sec:exp}

\vspace{-2eX}


We have extended the tool  \spen~\cite{SPENsite} with the proof strategy proposed in this paper. 
The entailments are written in an extension of the SMTLIB format used in the competition SL-COMP'14 for separation logic solvers.
It provides as output SAT, UNSAT or UNKNOWN, and a diagnosis for all these cases.

%
The solver starts with a normalization step, based on the boolean abstractions described in~\cite{EneaLSV14}, which saturates the input formulas with (dis)equalities between location variables implied by the semantics of separating conjunction. 
%
The entailments of data constraints are translated into satisfiability problems in the theory of integers with uninterpreted functions, discharged using an SMT solver dealing with this theory.

We have experimented the proposed approach  
on two sets of benchmarks\footnote{\url{http://www.liafa.univ-paris-diderot.fr/spen/benchmarks.html}}:

\vspace{-2eX}
\begin{description}
\item[RDBI:] verification conditions for proving the correctness of iterative procedures (delete, insert, search) over recursive data structures storing integer data:
sorted lists, 
binary search trees (BST),
AVL trees, and
red black trees (RBT).
\item[SL-COMP'14:] problems in the SL-COMP'14 benchmark, without data constraints, where the inductive definitions are syntactically compositional.
\end{description}
\vspace{-1eX}


\newcommand{\psearch}{\texttt{search}}
\newcommand{\pinsert}{\texttt{insert}}
\newcommand{\pdelete}{\texttt{delete}}

\begin{table}[t]
\caption{Experimental results on benchmark RDBI}
\vspace{-8mm}
\label{tab:exp-data}
\begin{center}
\setlength{\tabcolsep}{8pt}
{\footnotesize
\begin{tabular}{l|c|r|c|r|r|r}\hline
\textsf{Data structure} 
& \textsf{Procedure} & \textsf{\#VC} & \textsf{Lemma} & \textsf{$\limp_\mathbb{D}$} & \multicolumn{2}{c}{\textsf{Time (s)}} \\
&           &               & (\#b, \#r, \#p, \#c, \#d) &                          & \spen & SMT
\\\hline\hline
sorted lists 
  & \psearch & 4 & (1, 3, 3, 1, 3) & 5 & 1.108 & 0.10
 \\
  & \pinsert & 8 & (4, 6, 3, 1, 2) & 7 & 2.902 & 0.15
 \\
  & \pdelete & 4 & (2, 2, 4, 1, 1) & 6 & 1.108 & 0.10
\\\hline
BST
  & \psearch &  4 & (2, 3, 6, 2, 2) & 6 & 1.191 & 0.15
 \\
  & \pinsert & 14 & (15, 18, 27, 4, 6) & 19 & 3.911 & 0.55
 \\
  & \pdelete & 25 & (13, 19, 82, 8, 5) & 23 & 8.412 & 0.58
\\\hline
AVL 
  & \psearch &  4 & (2, 3, 6, 2, 2) & 6 & 1.573 & 0.15
 \\
  & \pinsert & 22 & (18, 28, 74, 6, 8) & 66 & 6.393 & 1.33 
\\\hline
RBT 
  & \psearch &  4 & (2, 3, 6, 2, 2) & 6 & 1.171 & 0.15
 \\
  & \pinsert & 21 & (27, 45, 101, 7, 10) & 80 & 6.962 & 2.53 
\\\hline
\end{tabular}}
\end{center}
\vspace{-6mm}
\end{table}
%
\noindent Tab.~\ref{tab:exp-data} provides the experiment results\footnote{The evaluations used a 2.53\,GHz Intel processor with 2\,GB, running Linux on VBox.}
for \textbf{RDBI}. 
The column \textsf{\#VC} gives the number of verification conditions considered for each procedure.
The column \textsf{Lemma} provides statistics about the lemma applications as follows:
\#b and \#r are the number of the applications of the lemmas corresponding to base resp. inductive rules,
\#c and \#d are the number of the applications of the composition resp. derived lemmas, and
\#p is the number of predicates matched syntactically, without applying lemmas.
Column \textsf{$\limp_\mathbb{D}$} gives the number of entailments between data constraints generated by \spen.
Column \textsf{Time-\spen} gives the ``system'' time spent by \spen\ on all verification conditions of a function\footnote{\spen\ does not implement a batch mode, each entailment is dealt separately, including the generation of lemma. The SMT solver is called on the files generated by \spen.}
excepting the time taken to solve the data constraints
by the SMT solver,
which is given in the column \textsf{Time-SMT}. 

Tab.~\ref{tab:exp-slcomp} provides a comparison of our approach (column \textsf{\spen}) with the decision procedure in \cite{EneaLSV14} (column \textsf{\spen-TA}) on the same set of benchmarks from SL-COMP'14. The times of the two decision procedures are almost the same, which demonstrates that our approach, as an extension of that in \cite{EneaLSV14}, is robust.


\begin{table}[t]
\caption{Experimental results on benchmark SL-COMP'14}
\vspace{-4eX}
\label{tab:exp-slcomp}
\setlength{\tabcolsep}{8pt}
\begin{center}
{\footnotesize
\begin{tabular}{l|r|c|r|r}\hline
\textsf{Data structure}
 & \textsf{\#VC} & \textsf{Lemma} & \multicolumn{2}{c}{\textsf{Time-\spen (s)}} \\
 &          & (\#b, \#r, \#p, \#c, \#d) & \spen & \spen-TA 
\\\hline\hline
Nested linked lists 
& 16 & (17,47,14,8,0) & 4.428 & 4.382
\\\hline
Skip lists 2 levels 
& 4 & (11,16,1,1,0) & 1.629 & 1.636
\\\hline
Skip lists 3 levels
& 10 & (16,32,29,17,0) & 3.858 & 3.485
\\\hline
\end{tabular}}
\end{center}
\vspace{-4eX}
\end{table}

%



\vspace{-4mm}
\section{Related work}
\vspace{-2eX}

There have been many works on the verification of programs manipulating mutable data structures in general and the use of separation logic, e.g.,~\cite{AHJ+13,AGH+14,BCO05,BDP11,BPZ05,CDN+12,CHL11,CHO+11,EneaLSV14,GVA07,IBI+14,IRS13,IRV14,NC08,PWZ14,RBH07,ZKR08,ChuJT14}. 
In the following, we discuss those which are closer to our approach.

The prover SLEEK~\cite{CDN+12,NC08} provides proof strategies for proving entailments of SL formulas. 
These strategies are also based on lemmas, relating inductive definitions, but differently from our approach, these lemmas are supposed to be given by the user (SLEEK can prove the correctness of the lemmas once they are provided). Our approach is able to discover and synthesize the lemmas systematically, efficiently, and automatically. 

The natural proof approach DRYAD~\cite{QGS+13,Pek:2014:NPD:2666356.2594325} can prove automatically the correctness of programs against the specifications given by separation logic formulas with inductive definitions. Nevertheless, the lemmas are still supposed to be provided by the users in DRYAD, while our approach can generate the lemmas automatically. Moreover, DRYAD does not provide an independent solver to decide the entailment of separation logic formulas, which makes difficult to compare the performance of our tool with that of DRYAD. In addition, the inductive definitions used in our paper enable succinct lemmas, far less complex than those used in DRYAD, which include complex constraints on data variables and the magic wand. 

The method of cyclic proofs introduced by~\cite{BDP11} and extended recently in~\cite{ChuJT14} proves the entailment of two SL formulas 
by using induction on the paths of proof trees.
They are not generating the lemma, but the method is able to (soundly) check intricate lemma given by the user, even ones which are out of the scope of our method, e.g., lemmas concerning the predicate $RList$ which is defined by unfolding the list segments from the end, instead of the beginning.
The cyclic proofs method can be seen like a dynamic lemma generation using complex reasoning on proof trees, while our method generates lemma statically by simple checks on the inductive definitions.
We think that our lemma generator could be used in the cyclic proof method to cut proof trees.

The tool SLIDE~\cite{IRS13,IRV14} provides decision procedures for fragments of SL based on reductions to the language inclusion problem of tree automata. Their fragments contain no data or size constraints. In addition, the EXPTIME lower bound complexity is an important obstacle for scalability. Our previous work~\cite{EneaLSV14} introduces a decision procedure based on reductions to the membership problem of tree automata which however is not capable of dealing with data constraints.

The tool GRASShopper~\cite{PWZ14} 
is based on translations of SL fragments to first-order logic with reachability predicates, and the use of SMT solvers to deal with the latter. The advantage is the  integration with other SMT theories to reason about data. However, this approach considers a limited class of inductive definitions (for linked lists and trees) and is incapable of dealing with the size or multiset constraints, thus unable to reason about AVL or red-black trees. 




The truncation point approach~\cite{GVA07} provides a method to specify and verify programs based on separation logic with inductive definitions that may specify truncated data structures with multiple holes, but it cannot deal with data constraints. Our approach can also be extended to cover such inductive definitions.


\vspace{-4mm}
\section{Conclusion}
\vspace{-3mm}

We proposed a novel approach for automating program proofs based on Separation Logic with inductive definitions. 
This approach consists of 
(1) efficiently checkable syntactic criteria for recognizing inductive definitions that satisfy crucial lemmas in such proofs and 
(2) a novel proof strategy for applying these lemmas.  
The proof strategy relies on syntactic matching of spatial atoms 
and on SMT solvers for checking data constraints. 
We have implemented this approach in our solver \spen\ and applied it successfully to a representative set of examples, coming from iterative procedures for binary search trees or lists.  

In the future, we plan to investigate extensions to 
more general inductive definitions  
by investigating ideas from~\cite{QGS+13,ChuJT14} to extend our proof strategy.  
From a practical point of view, apart from improving the implementation of our proof strategy, we plan to integrate it into the program analysis framework {\sf Celia}~\cite{celia}.

\bibliographystyle{abbrv}

\bibliography{sl}

\newpage

\appendix

\begin{appendix}


\section{Proofs in Sec.~\ref{sec:comp}}

\noindent {\bf Theorem \ref{thm-one-pred-compos}}.
\emph{Suppose that $\Pp$ is a set of inductively defined predicates. If $P \in \Pp$ is syntactically compositional, then $P$ is semantically compositional. }

\begin{proof}
Suppose $P$ is syntactically compositional and has parameters $(\orar{\alpha},\orar{\beta},\orar{\xi}\ )$.

It is sufficient to prove the following claim. 
\begin{quote}
For each pair $(s,h)$, if $(s,h) \models P(\orar{\alpha}_1,\orar{\alpha}_2,\orar{\xi'}) \sep P(\orar{\alpha}_2,\orar{\alpha}_3,\orar{\xi'})$, then $(s,h) \models P(\orar{\alpha}_1,\orar{\alpha}_3,\orar{\xi'})$.
\end{quote}

We prove the claim by induction on the size of $\ldom(h)$.

Suppose for each $i: 1 \le i \le 3$, $\orar{\alpha}_i= E_i\ v_i$, where $E_i$ and $v_i$ are respectively location and data variables.

Since $(s,h) \models P(\orar{\alpha}_1,\orar{\alpha}_2,\orar{\xi'}) \sep P(\orar{\alpha}_2,\orar{\alpha}_3,\orar{\xi'})$, there are $h_1,h_2$ such that $h=h_1 \sep h_2$, $(s,h_1) \models P(\orar{\alpha}_1,\orar{\alpha}_2,\orar{\xi'})$, and $(s,h_2) \models P(\orar{\alpha}_2,\orar{\alpha}_3,\orar{\xi'})$. 

If $(s,h_1) \models \bigwedge \limits_{i=1}^2 \alpha_{1,i} = \alpha_{2,i} \land \slemp$, then $\ldom(h_1)=\emptyset$, and $h_2=h$. From this, we deduce that $(s,h) \models P(\orar{\alpha}_1,\orar{\alpha}_3,\orar{\xi'})$.

Otherwise, there are a recursive rule of $P$, say $P(\orar{\alpha},\orar{\beta},\orar{\xi}\ )::=\exists \orar{X}.\ \Pi \land \Sigma_1 \sep \Sigma_2 \sep P(\orar{\gamma},\orar{\beta},\orar{\xi}\ )$, and an extension of $s$, say $s'$, such that $(s',h_1) \models \Pi' \land \Sigma'_1 \sep \Sigma'_2 \sep P(\orar{\gamma'},\orar{\alpha}_2,\orar{\xi'})$, where $\Pi', \Sigma'_1, \Sigma'_2, \gamma'$ are obtained from $\Pi,\Sigma_1,\Sigma_2,\gamma$ by replacing $\orar{\alpha}, \orar{\beta}, \orar{\xi}$ with $\orar{\alpha_1}, \orar{\alpha}_2,\orar{\xi'}$ respectively. 
From this, we deduce that there are $h_{1,1},h_{1,2},h_{1,3}$ such that $h_1=h_{1,1} \sep h_{1,2} \sep h_{1,3}$, $(s',h_{1,1}) \models \Sigma'_1$, $(s',h_{1,2}) \models \Sigma'_2$, and $(s',h_{1,3}) \models P(\orar{\gamma'},\orar{\alpha}_2,\orar{\xi'})$.  
Then $(s',h_{1,3} \sep h_2) \models P(\orar{\gamma'},\orar{\alpha}_2,\orar{\xi'}) \sep P(\orar{\alpha}_2,\orar{\alpha}_3, \orar{\xi'})$. 
From the induction hypothesis, we deduce that $(s',h_{1,3} \sep h_2) \models P(\orar{\gamma'},\orar{\alpha}_3, \orar{\xi'})$. 
Then $(s',h_{1,1} \sep h_{1,2} \sep h_{1,3} \sep h_2) \models \Pi' \land \Sigma'_1 \sep \Sigma'_2 \sep P(\orar{\gamma'},\orar{\alpha}_3, \orar{\xi'})$. We then deduce that $(s,h) \models \exists \orar{X}.\ \Pi' \land \Sigma'_1 \sep \Sigma'_2 \sep P(\orar{\gamma'},\orar{\alpha}_3, \orar{\xi'})$.  

To prove $(s,h) \models P(\orar{\alpha_1},\orar{\alpha}_3,\orar{\xi'})$, it is sufficient to prove that $(s,h) \models \exists \orar{X}.\ \Pi'' \land \Sigma''_1 \sep \Sigma''_2 \sep P(\orar{\gamma''},\orar{\alpha}_3,\orar{\xi'})$, where $\Pi'', \Sigma''_1,\Sigma''_2,\orar{\gamma''}$ are obtained from $\Pi,\Sigma_1,\Sigma_2,\orar{\gamma}$ by replacing $\orar{\alpha}, \orar{\beta}, \orar{\xi}$ with $\orar{\alpha}_1, \orar{\alpha}_3, \orar{\xi'}$ respectively. 

From the fact that no variables from $\orar{\beta}$ occur in $\Pi$, $\Sigma_1$, $\Sigma_2$, or $\orar{\gamma}$, we know that $\Pi''=\Pi'$, $\Sigma''_1=\Sigma'_1$, $\Sigma''_2 = \Sigma'_2$, and $\orar{\gamma''}=\orar{\gamma'}$. 
Since $(s,h) \models \exists \orar{X}.\ \Pi' \land \Sigma'_1 \sep \Sigma'_2 \sep P(\orar{\gamma'},\orar{\alpha}_3, \orar{\xi'})$, we have already proved that $(s,h) \models \exists \orar{X}.\ \Pi'' \land \Sigma''_1 \sep \Sigma''_2 \sep P(\orar{\gamma''},\orar{\alpha}_3,\orar{\xi'})$. The proof is done.
\qed
\end{proof} 

\section{Proofs in Sec.~\ref{sec:derived}}

\noindent {\bf Theorem \ref{thm-completion}}.
\emph{Let $P \in \Pp$ be a syntactically compositional predicate with the parameters $(\orar{\alpha},\orar{\beta},\orar{\xi})$, and $P' \in \Pp$ with the parameters $(\orar{\alpha},\orar{\xi}\ )$. If $P'$ is a completion of $P$ with respect to $\orar{c}$, then $P'(\orar{\alpha},\orar{\xi}\ ) \Leftrightarrow P(\orar{\alpha},\orar{c},\orar{\xi}\ )$ and  $\exists \vec{\beta}. \  P(\orar{\alpha},\orar{\beta},\orar{\xi}\ ) \sep P'(\orar{\beta},\orar{\xi}\ ) \Rightarrow P'(\orar{\alpha}, \orar{\xi}\ )$ hold}.

\begin{proof}

The fact $P'(\orar{\alpha},\orar{\xi}) \Leftrightarrow P(\orar{\alpha},\orar{c},\orar{\xi})$ can be proved easily by an induction on the size of the domain of the heap structures.

The argument for $\exists \vec{\beta}.\ P(\orar{\alpha},\orar{\beta},\orar{\xi}) \sep P'(\orar{\beta},\orar{\xi}) \Rightarrow P'(\orar{\alpha}, \orar{\xi})$ goes as follows:  Suppose $(s,h) \models P(\orar{\alpha},\orar{\beta},\orar{\xi}) \sep P'(\orar{\beta},\orar{\xi})$. Then there are $h_1,h_2$ such that $h=h_1 \sep h_2$, $(s,h_1) \models P(\orar{\alpha},\orar{\beta},\orar{\xi})$, and $(s,h_2) \models P'(\orar{\beta},\orar{\xi})$. From the fact that $P'(\orar{\beta},\orar{\xi}) \Leftrightarrow P(\orar{\beta},\orar{c},\orar{\xi})$, we know that $(s,h_2) \models P(\orar{\beta},\orar{c},\orar{\xi})$. Therefore, $(s,h) \models P(\orar{\alpha},\orar{\beta},\orar{\xi}) \sep P(\orar{\beta},\orar{c}, \orar{\xi})$. From Theorem \ref{thm-one-pred-compos}, we deduce that $(s,h) \models P(\orar{\alpha},\orar{c}, \orar{\xi})$. From the fact $P(\orar{\alpha},\orar{c}, \orar{\xi}) \Leftrightarrow P'(\orar{\alpha}, \orar{\xi})$, we conclude that $(s,h) \models P'(\orar{\alpha}, \orar{\xi})$.
\qed
\end{proof}

\noindent {\bf Theorem \ref{thm-strong}}.
\emph{Let $P,P' \in \Pp$ be two syntactically compositional inductively defined predicates with the same set of parameters $(\orar{\alpha},\orar{\beta},\orar{\xi}\ )$. If $P'$ is stronger than $P$, then the entailment $P'(\orar{\alpha},\orar{\beta},\orar{\xi}\ )\Rightarrow P(\orar{\alpha},\orar{\beta},\orar{\xi}\ )$ and $\exists \vec{\beta}. \ P'(\orar{\alpha},\orar{\beta},\orar{\xi}\ ) \sep P(\orar{\beta},\orar{\gamma},\orar{\xi}\ ) \Rightarrow P(\orar{\alpha},\orar{\gamma},\orar{\xi}\ )$ hold.}

\begin{proof}
We first show that $P'(\orar{\alpha},\orar{\beta},\orar{\xi}\ )\Rightarrow P(\orar{\alpha},\orar{\beta},\orar{\xi}\ )$. By induction on the size of $\ldom(h)$, we prove the following fact: For each $(s,h)$, if $(s,h) \models P'(\orar{\alpha},\orar{\beta},\orar{\xi}\ )$, then $(s,h) \models P(\orar{\alpha},\orar{\beta},\orar{\xi}\ )$.

Suppose $(s,h) \models P'(\orar{\alpha},\orar{\beta},\orar{\xi}\ )$. 

If $(s,h) \models \bigwedge \limits_{i=1}^2 \alpha_i = \beta_i \land \slemp$, since $P'$ and $P$ have the same base rule, we deduce that $(s,h) \models P(\orar{\alpha},\orar{\beta},\orar{\xi}\ )$. 

Otherwise, there are a recursive rule of $P'$, say $P'(\orar{\alpha},\orar{\beta},\orar{\xi}\ )::= \exists \orar{X}.\ \Pi' \land \Sigma_1 \sep \Sigma_2 \sep P'(\orar{\gamma},\orar{\beta}, \orar{\xi}\ )$, and an extension of $s$, say $s'$, such that $(s',h) \models \Pi' \land \Sigma_1 \sep \Sigma_2 \sep P'(\orar{\gamma},\orar{\beta},\orar{\xi}\ )$. Then there are $h_1,h_2,h_3$ such that $h=h_1 \sep h_2 \sep h_3$, $(s', h_1) \models \Sigma_1$, $(s', h_2) \models \Sigma_2$, and $(s', h_3) \models P'(\orar{\gamma},\orar{\beta},\orar{\xi}\ )$.  From the induction hypothesis, we deduce that $(s', h_3) \models P(\orar{\gamma},\orar{\beta},\orar{\xi}\ )$.
Moreover, from the assumption, we know that there is a recursive rule of $P$  of the form $P(\orar{\alpha},\orar{\beta},\orar{\xi}\ )::= \exists \orar{X}.\ \Pi \land \Sigma_1 \sep \Sigma_2 \sep P(\orar{\gamma},\orar{\beta}, \orar{\xi}\ )$, such that $\Pi' \Rightarrow \Pi$ holds. Then it follows that $(s',h_1 \sep h_2 \sep h_3) \models \Pi \land \Sigma_1 \sep \Sigma_2 \sep P(\orar{\gamma},\orar{\beta},\orar{\xi}\ )$. We then deduce that $(s,h) \models \exists \orar{X}.\ \Pi \land \Sigma_1 \sep \Sigma_2 \sep P(\orar{\gamma},\orar{\beta},\orar{\xi}\ )$. From this, we conclude that $(s,h) \models P(\orar{\alpha},\orar{\beta},\orar{\xi}\ )$.

We then prove the second claim of the theorem.

From the argument above, we know that $P'(\orar{\alpha},\orar{\beta},\orar{\xi}\ ) \Rightarrow P(\orar{\alpha},\orar{\beta},\orar{\xi}\ )$ holds. Then $P'(\orar{\alpha},\orar{\beta},\orar{\xi}\ ) \sep P(\orar{\beta},\orar{\gamma},\orar{\xi}\ ) \Rightarrow P(\orar{\alpha},\orar{\beta},\orar{\xi}\ ) \sep P(\orar{\beta},\orar{\gamma},\orar{\xi}\ )$ holds. In addition, from Theorem \ref{thm-one-pred-compos}, we know that $P(\orar{\alpha},\orar{\beta},\orar{\xi}\ ) \sep P(\orar{\beta},\orar{\gamma},\orar{\xi}\ ) \Rightarrow P(\orar{\alpha},\orar{\gamma},\orar{\xi}\ )$ holds. Therefore,we conclude that $P'(\orar{\alpha},\orar{\beta},\orar{\xi}\ ) \sep P(\orar{\beta},\orar{\gamma},\orar{\xi}\ ) \Rightarrow P(\orar{\alpha},\orar{\gamma},\orar{\xi}\ )$.
\qed
\end{proof}

\noindent{\bf Theorem \ref{thm-border-contr}}.
\emph{Let $P \in \Pp$ be a syntactically compositional predicate with the parameters $(\orar{\alpha}, \orar{\beta}, \orar{\xi}\,)$ and $P' \in \Pp$ be an inductive predicate with the parameters $(\orar{\alpha}, \orar{\beta}, \orar{\xi'})$. If $P'$ is a static-parameter contraction of $P$ with the contraction function $\eta$, then $P'(\orar{\alpha}, \orar{\beta}, \orar{\xi'}) \Leftrightarrow P(\orar{\alpha}, \orar{\beta}, \eta(\vec{\xi}\ ))$ and $\exists\orar{\beta}.\ P(\orar{\alpha}, \orar{\beta}, \eta(\vec{\xi}\ )) \sep P'(\orar{\beta}, \orar{\gamma}, \orar{\xi'}) \Rightarrow P'(\orar{\alpha}, \orar{\gamma}, \orar{\xi'})$ hold.}

\begin{proof}
The first claim can be proved by induction on the size of the domain of the heap structures.

The argument for the second claim goes as follows: From the fact that $P'(\orar{\beta}, \orar{\gamma}, \orar{\xi'}) \Leftrightarrow P(\orar{\beta}, \orar{\gamma}, \eta(\vec{\xi}\ ))$, we deduce that
\[
\exists\orar{\beta}.\ P(\orar{\alpha}, \orar{\beta}, \eta(\vec{\xi}\ )) \sep P'(\orar{\beta}, \orar{\gamma}, \orar{\xi'}) \ \ \Rightarrow 
\ \ P(\orar{\alpha}, \orar{\beta}, \eta(\vec{\xi}\ )) \sep P(\orar{\beta}, \orar{\gamma}, \eta(\vec{\xi}\ )).
\]

From Theorem \ref{thm-one-pred-compos}, we know that
\[P(\orar{\alpha}, \orar{\beta}, \eta(\vec{\xi}\ )) \sep P(\orar{\beta}, \orar{\gamma}, \eta(\vec{\xi}\ )) \Rightarrow 
P(\orar{\alpha}, \orar{\gamma}, \eta(\vec{\xi}\ )).\]
Then the second claim follows from the fact $P(\orar{\alpha}, \orar{\gamma}, \eta(\orar{\xi}\ ) ) \Leftrightarrow P'(\orar{\alpha}, \orar{\gamma}, \orar{\xi'})$. \qed
\end{proof}

\section{Extensions of the lemmas}\label{app:extensions}

In this section, we discuss how the the basic idea of syntactical compositionality can be extended in various ways.

\subsection{Multiple location and data parameters} 

At first, we would like to emphasize that although we restrict our discussions on compositional predicates $P(\orar{\alpha}, \orar{\beta}, \orar{\xi})$ to the special case that $\orar{\alpha}$ (resp. $\orar{\beta}$) contain only two parameters: one location parameter, and one data parameter. But all the results about the lemmas can be generalized smoothly to the situation that $\orar{\alpha}$ and $\orar{\beta}$ contain multiple location and data parameters.

\subsection{Pseudo-composition lemmas}

We then consider syntactically pseudo-compositional predicates.

We still use the binary search trees to illustrate the idea.

Suppose $neqbsthole$ is the predicate defined by the same rules as $bsthole$, with the modification that $E \neq F$ is added to the body of each inductive rule. Then $\it neqbsthole$ is not syntactically compositional anymore and the composition lemma 
\[
\begin{array}{l}
\exists E_2,M_2. \ neqbsthole(E_1,M_1,E_2,M_2) \sep neqbsthole(E_2,M_2,E_3,M_3) \Rightarrow\\
\hspace{7cm} neqbsthole(E_1,M_1,E_3,M_3)
\end{array}
\]
\noindent does not hold. This is explained as follows: Suppose $h= h_1 \sep h_2$ (where $h=h_1 \sep h_2$ denotes that $h_1$ and $h_2$ are domain disjoint and $h$  is the union of $h_1$ and $h_2$), $(s,h_1) \models neqbsthole(E_1,M_1,E_2,M_2)$ and $(s,h_2) \models neqbsthole(E_2,M_2,E_3,M_3)$, in addition, both $\ldom(h_1)$ and $\ldom(h_2)$ are nonempty. Then from the inductive definition of $neqbsthole$, we deduce that $s(E_1) \neq s(E_2)$ and $s(E_2) \neq s(E_3)$. On the other hand, $(s,h) \models bsthole1(E_1,M_1,E_3,M_3)$ requires that $s(E_1) \neq s(E_3)$, which cannot be inferred from $s(E_1) \neq s(E_2)$ and $s(E_2) \neq s(E_3)$ in general. Nevertheless, the entailment 
\noindent 
\[\begin{array}{l}
\exists E_2,M_2.\ neqbsthole(E_1,M_1,E_2,M_2) \sep neqbsthole(E_2,M_2,E_3,M_3) \ \sep \\
\hspace{1.5cm} E_3 \mapsto ((left,X),(right,Y),(data,v)) \Rightarrow \\
\hspace{1.5cm} neqbsthole(E_1,M_1,E_3,M_3) \sep E_3 \mapsto ((left,X),(right,Y),(data,v))
\end{array}
\]
\noindent holds since the information $E_1 \neq E_3$ can be inferred from the fact that $E_3$ is allocated and separated from $E_1$. Therefore, intuitively, in this situation, the composition lemma can be applied under the condition that we already know that $E_1 \neq E_3$. We call this as pseudo-compositionality. Our decision procedure can be generalized to apply the pseudo-composition lemmas when proving the entailment of two formulas.

\subsection{Data structures with parent pointers}

Next, we show how our ideas can be generalized to the data structures with parent pointers, e.g. doubly linked lists or trees with parent pointers. We use binary search trees with parent pointers to illustrate the idea. We can define the predicates $prtbst(E,Pr,M)$ and $prtbsthole(E, Pr_1, M_1, F, Pr_2, M_2)$ to describe respectively binary search trees with parent pointers and binary search trees with parent pointers and one hole. The intuition of $E,F$ are still the source and the hole, while $Pr$ and $Pr_1$ (resp. $Pr_2$) are the parent of $E$ (resp. $F$) (the definition of $prtbst$ is omitted here).
\[
\begin{array}{l l} 
prtbsthole(E, Pr_1, M_1,F, Pr_2, M_2) &::= E = F \land \slemp \land Pr_1 = Pr_2 \land M_1 = M_2  \\[1mm]
prtbsthole(E, Pr_1, M_1,F, Pr_2, M_2) &::=  \exists X,Y,M_3,M_4,v.\\
& \hspace{-1.5cm} \ E \mapsto \{(left,X), (right,Y),(parent,Pr_1), (data,v)\} \\
& \hspace{-1.5cm} \sep\ prtbst(X,E,M_3) \sep prtbsthole(Y,E, M_4, F, Pr_2, M_2) \\
& \hspace{-1.5cm} \land \ M_1=\{v\} \cup M_3 \cup M_4 \land M_3 < v < M_4 \\ [1mm]
prtbsthole(E, Pr_1, M_1,F, Pr_2, M_2)  &::= \exists X,Y,M_3,M_4,v.\\
& \hspace{-1.5cm}  E \mapsto \{(left,X), (right,Y),(parent,Pr_1), (data,v)\}\\
& \hspace{-1.5cm} \sep \ prtbsthole(X,E,M_3, F, Pr_2, M_2) \sep prtbst(Y,E, M_4) \\
& \hspace{-1.5cm}  \land \ M_1=\{v\} \cup M_3 \cup M_4 \land M_3 < v < M_4
\end{array}
\]
Then the predicate $prtbsthole$ enjoys the composition lemma
\[\begin{array} {l}
\exists E_2, Pr_2, M_2. \ prtbsthole(E_1, Pr_1, M_1,E_2, Pr_2, M_2) \ \sep \\
\hspace{2.3cm} prtbsthole(E_2, Pr_2, M_2,E_3, Pr_3, M_3) \Rightarrow \\
\hspace{6cm} prtbsthole(E_1, Pr_1, M_1,E_3, Pr_3, M_3).
\end{array}
\]
%

%
%
%

\subsection{Points-to atom in base rules}

Finally, we discuss the constraint that the base rule of a syntactically compositional predicate has an empty spatial atom. We use the  predicates $lsegeven$ and $lsegodd$ to illustrate the idea. 
\[\begin{array}{ll}
lsegeven(E,F)& ::= E=F \land \slemp,\\[1mm]
lsegeven(E,F) & ::= \exists X,Y. \ E \mapsto (next,X) \sep X \mapsto (next,Y) \sep lsegeven(Y,F).
\end{array}
\]
The definition of $lsegodd(E,F)$ can be obtained from that of $lsegeven(E,F)$ by replacing the base rule with the rule $lsegodd(E,F) ::= E \mapsto (next,F)$. 
The only difference between the inductive definition of $lsegeven$ and and that of $lsegodd$ is that $lsegeven$ has an empty base rule, while $lsegodd$ does not. From this, we deduce that 
\[lsegodd(E,F) \Leftrightarrow \exists X.\ E \mapsto \{(next,X)\} \sep lsegeven(X,F).\]
This idea can be generalized to arbitrary syntactically compositional predicates.


\section{Full example of Sec.~\ref{sec:slice}}
\label{sec:app-ex}

We provide here the full details of the example considered in Section~\ref{sec:slice}.

Consider the following entailment which states that two cells linked by the $\fnext$ pointer field, and storing ordered data values, form a sorted list segment:
\vspace{-1eX}
{\small
\begin{align*}
\varphi_1 & ::= x_1\neq\nil\land x_2\neq \nil \land v_1 < v_2\land x_1 \mapsto \{(\fnext,x_2),(\fdata,v_1)\} \\
& \hspace{34eX}\sep\ x_2 \mapsto \{(\fnext,\nil), (\fdata,v_2)\} \\
\varphi_2 & ::= \exists M.\ \plseg(x_1, M, \nil,\emptyset) \land v_2\in M,
\end{align*}
}

\vspace{-5mm}
\noindent
where $\plseg$ has been defined in Sec.~\ref{sec:intro} (eq.~(\ref{rule:lseg-base})--(\ref{rule:lseg-rec})).  

For convenience, let 
\vspace{-1eX}
\begin{align*}
\Pi_1 & ::= x_1\neq\nil\land x_2\neq \nil \land v_1 < v_2, \\
\Sigma_1 & ::=  x_1 \mapsto \{(\fnext,x_2),(\fdata,v_1)\} \sep x_2 \mapsto \{(\fnext,\nil), (\fdata,v_2)\},\\ 
\Pi_2 & ::=  v_2\in M, \\
\Sigma_2 & ::= \plseg(x_1, M, \nil,\emptyset).
\end{align*}

\noindent {\bf The first application of the rule} \textsc{Slice}. Since the right-hand side contains a single spatial atom,  the rule $\textsc{Slice}$ generates a sub-goal $\Pi_1 \land \Sigma_1 \models_{\eta} \exists M.\ \Sigma_2$. For the sub-goal, the syntactic matching (rule \textsc{Match1}) cannot be applied. Instead, we apply the rule $\textsc{Lemma}$ using a lemma $L$ that corresponds to the inductive rule of $\plseg$,
i.e., eq.~(\ref{rule:lseg-rec}) (page~\pageref{rule:lseg-rec}): 
\[
\begin{array}{l}
L::=\exists X,M_1,v.\ x_1 \mapsto \{(\fnext,X), (\fdata,v)\} \sep \plseg(X,M_1,\nil,\emptyset) 
\land \\
\hspace{4.5cm} M=\{v\}\cup M_1 \land v \le M_1 \limp \plseg(x_1, M, \nil,\emptyset).
\end{array}
\]
For convenience, let 
\vspace{-1eX}
\begin{align*}
\Pi & ::= M=\{v\}\cup M_1 \land v \le M_1, \\
\Sigma & ::=   x_1 \mapsto \{(\fnext,X), (\fdata,v)\} \sep \plseg(X,M_1,\nil,\emptyset). 
\end{align*}

\noindent {\bf The first application of the rule} \textsc{Lemma}. Since $root(L)::=x_1 \mapsto \{(\fnext,X), (\fdata,v)\}$, the rule \textsc{Lemma} generates a sub-goal $\Pi_1 \land \Sigma'_1 \models^{SUB}_{\eta_1} \exists X,v.\ root(L)$, where $\Sigma'_1::=x_1 \mapsto \{(\fnext,x_2),(\fdata,v_1)\}$. Then the rule $\textsc{Match1}$ is applied, resulting in a quantifier instantiation $\eta_1=\{X \rightarrow x_2, v \rightarrow v_1\}$. Note that, since $\eqmap(\eta_1|_{\freev(\exists X,v. root(L))})::= \true$, the entailment $\Pi_1 \land \eqmap(\eta_1|_{\{X,v\}}) \models \eqmap(\eta_1|_{\freev(\exists X,v. root(L))})$ holds. 
The variable substitution $\eta_1$ is used to instantiate the existentially quantified variables in the remaining part of the lemma, that is, $\Pi \land  \plseg(X,M_1,\nil,\emptyset)$, resulting into the formula 
\[
\begin{array}{l c l}
\eta_1(\Pi \land \plseg(X,M_1,\nil,\emptyset)) & ::= & \\
& & \hspace{-2.5cm} M=\{v_1\} \cup M_1 \land v_1 \le M_1 \land \plseg(x_2,M_1,\nil,\emptyset).
\end{array}
\] 
Then, the rule $\textsc{Lemma}$ generates another sub-goal $\Pi_1 \land \Sigma''_1 \models_{\eta_2} \exists M,M_1.\ \eta_1(\Pi \land \plseg(X,M_1,\nil,\emptyset))$, where $\Sigma''_1 ::= x_2 \mapsto \{(\fnext,\nil), (\fdata,v_2)\}$.

\smallskip
\noindent {\bf The second application of the rule} \textsc{Slice}. For the sub-goal $\Pi_1 \land \Sigma''_1 \models_{\eta_2} \exists M,M_1.\ \eta_1(\Pi \land \plseg(X,M_1,\nil,\emptyset))$, the rule $\textsc{Slice}$ is applied again. Since there is a single spatial atom in the RHS, the rule $\textsc{Slice}$ generates a sub-goal $\Pi_1 \land \Sigma''_1 \models_{\eta_3} \exists M_1.\ \plseg(x_2,M_1,\nil,\emptyset)$.

\smallskip

\noindent {\bf The second application of the rule} \textsc{Lemma}. For the sub-goal $\Pi_1 \land \Sigma''_1 \models_{\eta_3} \exists M_1.\ \plseg(x_2,M_1,\nil,\emptyset)$, the rule \textsc{Lemma} is applied again, using the  lemma $L'$ (still corresponding to the inductive rule of $\plseg$),
\[
\begin{array}{l}
L'::=\exists X',M'_1,v'.\ x_2 \mapsto \{(\fnext,X'), (\fdata,v')\} \sep \plseg(X', M'_1,\nil,\emptyset) 
\land \\
\hspace{4.5cm} M_1=\{v'\}\cup M'_1 \land v' \le M'_1 \limp \plseg(x_2, M_1, \nil,\emptyset).
\end{array}
\]
For convenience, let 
\vspace{-1eX}
\begin{align*}
\Pi' & ::= M_1=\{v'\}\cup M'_1 \land v' \le M'_1, \\
\Sigma' & ::=   x_2 \mapsto \{(\fnext,X'), (\fdata,v')\} \sep \plseg(X', M'_1,\nil,\emptyset). 
\end{align*}
Since $root(L')::= x_2 \mapsto \{(\fnext,X'), (\fdata,v')\}$, the rule \textsc{Lemma} generates a sub-goal $\Pi_1 \land \Sigma''_1 \models^{SUB}_{\eta'_1} \exists X',v'.\ root(L')$. Then the rule $\textsc{Match1}$ is applied, resulting in a quantifier instantiation $\eta'_1=\{X' \rightarrow \nil, v' \rightarrow v_2\}$. Note that, since $\eqmap(\eta'_1|_{\freev(\exists X',v'. root(L'))})::= \true$, the entailment $\Pi_1 \land \eqmap(\eta'_1|_{\{X',v'\}}) \models \eqmap(\eta'_1|_{\freev(\exists X',v'. root(L'))})$ holds. 
The variable substitution $\eta'_1$ is used to instantiate the existentially quantified variables in the remaining part of the lemma, that is, $\Pi' \land \plseg(X',M'_1,\nil,\emptyset)$, resulting into the formula 
\[
\begin{array}{l c l}
\eta'_1(\Pi' \land \plseg(X',M'_1,\nil,\emptyset)) & ::= & \\
& & \hspace{-1.5cm} M_1=\{v_2\}\cup M'_1 \land v_2 \le M'_1 \land \plseg(\nil, M'_1,\nil,\emptyset).
\end{array}
\] 
 Then, the rule $\textsc{Lemma}$ generates another sub-goal $\Pi_1 \land \slemp \models_{\eta'_2} \exists M_1, M'_1.\ \eta'_1(\Pi' \land \plseg(X',M'_1,\nil,\emptyset))$. 

\smallskip

\noindent {\bf The third application of the rule} \textsc{Slice}.  For the sub-goal $\Pi_1 \land \slemp \models_{\eta'_2} \exists M_1,M'_1.\ \eta'_1(\Pi' \land \plseg(X',M'_1,\nil,\emptyset))$, since there is only one spatial atom $\plseg(\nil, M'_1,\nil,\emptyset)$ in the RHS, the rule \textsc{Slice} generates a subgoal $\Pi_1 \land \slemp \models_ {\eta'_3} \exists M'_1.\ \plseg(\nil, M'_1,\nil,\emptyset)$, for which the rule \textsc{Lemma} is applied, with a lemma $L''$ corresponding to the base rule of $\plseg$, i.e., eq.~(\ref{rule:lseg-base}) (page~\pageref{rule:lseg-rec}), 
\[L'' \hspace{2mm}  ::=\hspace{2mm} \nil = \nil \land \slemp \land M'_1= \emptyset \hspace{2mm} \Rightarrow \hspace{2mm} \plseg(\nil, M'_1,\nil,\emptyset).\]
 
\noindent {\bf The third application of the rule} \textsc{Lemma}.  Since $root(L'')::= \slemp$, the rule \textsc{Lemma} generates a sub-goal $\Pi_1 \land \slemp \models^{SUB}_{\eta''_1} root(L'')$. Then the rule $\textsc{Match1}$ is applied, resulting in a quantifier instantiation $\eta''_1=\emptyset$. Note that, since $\eqmap(\eta''_1|_{\freev(root(L''))})::= \true$, the entailment $\Pi_1 \land \eqmap(\eta''_1|_{\emptyset}) \models \eqmap(\eta''_1|_{\freev(root(L''))})$ holds. 
The substitution $\eta''_1$ is used to instantiate the existential variables in the remaining part of the lemma, that is, $\nil = \nil \land M'_1= \emptyset$, resulting into the same formula. Then the rule \textsc{Lemma} generates a sub-goal $\Pi_1 \models_{\eta''_2} \exists M'_1.\ \nil = \nil \land M'_1= \emptyset$, which holds clearly with $\eta''_2=\emptyset$. Finally, $\eta''_1 \cup \eta''_2$ is extended with $\nil = \nil \land M'_1= \emptyset$, resulting into $\eta'_3=\{M'_1 \rightarrow \emptyset\}$.

\smallskip

\noindent {\bf The third application of the rule} \textsc{Slice} ({\bf continued}).  The variable substitution $\eta'_3$ is extended with $M_1=\{v_2\}\cup M'_1 \land v_2 \le M'_1$, resulting into $\eta'_2=\{M_1 \rightarrow \{v_2\} \cup \emptyset,\ M'_1 \rightarrow \emptyset\}$. Then the rule \textsc{Slice} generates a sub-goal $\Pi_1 \land \eqmap(\eta'_2) \models M_1=\{v_2\}\cup M'_1 \land v_2 \le M'_1$. Because $\eqmap(\eta'_2)=M_1 = \{v_2\} \cup \emptyset \land M'_1 = \emptyset$, we know that $\eqmap(\eta'_2) \models M_1=\{v_2\}\cup M'_1 \land v_2 \le M'_1$. Thus the sub-goal holds.

\smallskip
\noindent {\bf The second application of the rule} \textsc{Lemma} ({\bf continued}). The variable substitution $\eta'_1 \cup \eta'_2$ should be extended with $\Pi'::=M_1=\{v'\} \cup M'_1 \land v' \le M'_1$. Since $M_1 \in \dom(\eta'_2)$, the extension makes no effect. Then $\eta_3=(\eta'_1 \cup \eta'_2)|_{\{M_1\}}=\{M_1 \rightarrow \{v_2\} \cup \emptyset\}$.

\smallskip
\noindent {\bf The second application of the rule} \textsc{Slice} ({\bf continued}).  The variable substitution $\eta_3$ is extended with $M=\{v_1\} \cup M_1 \land v_1 \le M_1$, resulting into $\eta_2=\{M \rightarrow \{v_1\} \cup \{v_2\} \cup \emptyset, M_1 \rightarrow \{v_2\} \cup \emptyset\}$. Then the rule \textsc{Slice} generates a sub-goal $\Pi_1 \land \eqmap(\eta_2) \models M=\{v_1\} \cup M_1 \land v_1 \le M_1$. Since $\Pi_1::=x_1\neq\nil\land x_2\neq \nil \land v_1 < v_2$ and $\eqmap(\eta_2)::=M = \{v_1\} \cup \{v_2\} \cup \emptyset \land M_1 = \{v_2\} \cup \emptyset$, the sub-goal holds.

\smallskip
\noindent {\bf The first application of the rule} \textsc{Lemma} ({\bf continued}). The variable substitution $\eta_1 \cup \eta_2$ should be extended with $\Pi::=M=\{v\} \cup M_1 \land v \le M_1$. Since $M \in \dom(\eta_1 \cup \eta_2)$, this extension makes no effect. Then $\eta$ is obtained from $\eta_1 \cup \eta_2$ by restricting to $\{M\}$. So $\eta=\{M \rightarrow \{v_1\} \cup \{v_2\} \cup \emptyset\}$.

\smallskip
\noindent {\bf The first application of the rule} \textsc{Slice} ({\bf continued}). The variable substitution $\eta$ is extended with $\Pi_2::=v_2 \in M$, and still getting $\eta$. Finally, \textsc{Slice} generates the sub-goal $\Pi_1 \land \eqmap(\eta) \models \Pi_2$. Since $\eqmap(\eta)::=M = \{v_1\} \cup \{v_2\} \cup \emptyset$, the entailment $\eqmap(\eta) \models \Pi_2$ holds. Therefore, the sub-goal $\Pi_1 \land \eqmap(\eta) \models \Pi_2$ holds as well.

\end{appendix}

\end{document}